\documentclass[%
 reprint, 
 amsmath,amssymb,
 aps,
]{revtex4-2}

\usepackage{hyperref}
\usepackage{amsthm}
\usepackage{algorithm,algorithmic}

\usepackage[dvipsnames]{xcolor}

\usepackage{multirow}
\usepackage{booktabs}

\catcode`~=11 \def\UrlSpecials{\do\~{\kern -.15em\lower .7ex\hbox{~}\kern .04em}} \catcode`~=13

\allowdisplaybreaks[3]

\newcommand{\nn}{\nonumber}

\newcommand{\calC}{\mathcal{C}}

\newcommand{\calS}{\mathcal{S}}

\newcommand{\bA}{\mathbf{A}}

\newcommand{\bB}{\mathbf{B}}

\newcommand{\bD}{\mathbf{D}}

\newcommand{\bI}{\mathbf{I}}

\newcommand{\bM}{\mathbf{M}}

\newcommand{\bbP}{\mathbb{P}}

\DeclareMathAlphabet{\mathbsf}{OT1}{cmss}{bx}{n}
\DeclareMathAlphabet{\mathssf}{OT1}{cmss}{m}{sl}

\DeclareSymbolFont{bsfletters}{OT1}{cmss}{bx}{n}
\DeclareSymbolFont{ssfletters}{OT1}{cmss}{m}{n}
\DeclareMathSymbol{\bsfGamma}{0}{bsfletters}{'000}
\DeclareMathSymbol{\ssfGamma}{0}{ssfletters}{'000}
\DeclareMathSymbol{\bsfDelta}{0}{bsfletters}{'001}
\DeclareMathSymbol{\ssfDelta}{0}{ssfletters}{'001}
\DeclareMathSymbol{\bsfTheta}{0}{bsfletters}{'002}
\DeclareMathSymbol{\ssfTheta}{0}{ssfletters}{'002}
\DeclareMathSymbol{\bsfLambda}{0}{bsfletters}{'003}
\DeclareMathSymbol{\ssfLambda}{0}{ssfletters}{'003}
\DeclareMathSymbol{\bsfXi}{0}{bsfletters}{'004}
\DeclareMathSymbol{\ssfXi}{0}{ssfletters}{'004}
\DeclareMathSymbol{\bsfPi}{0}{bsfletters}{'005}
\DeclareMathSymbol{\ssfPi}{0}{ssfletters}{'005}
\DeclareMathSymbol{\bsfSigma}{0}{bsfletters}{'006}
\DeclareMathSymbol{\ssfSigma}{0}{ssfletters}{'006}
\DeclareMathSymbol{\bsfUpsilon}{0}{bsfletters}{'007}
\DeclareMathSymbol{\ssfUpsilon}{0}{ssfletters}{'007}
\DeclareMathSymbol{\bsfPhi}{0}{bsfletters}{'010}
\DeclareMathSymbol{\ssfPhi}{0}{ssfletters}{'010}
\DeclareMathSymbol{\bsfPsi}{0}{bsfletters}{'011}
\DeclareMathSymbol{\ssfPsi}{0}{ssfletters}{'011}
\DeclareMathSymbol{\bsfOmega}{0}{bsfletters}{'012}
\DeclareMathSymbol{\ssfOmega}{0}{ssfletters}{'012}

\DeclareMathOperator*{\argmax}{arg\,max}

\newcommand{\Bin}{\mathrm{Bin}}
\newcommand{\Normal}{\mathcal{N}}

\newcommand{\bzero}{\mathbf{0}}

\newtheorem{assumption}{Assumption}
\newtheorem{data model}{Data Model}

\newcommand{\qednew}{\nobreak \ifvmode \relax \else
      \ifdim\lastskip<1.5em \hskip-\lastskip
      \hskip1.5em plus0em minus0.5em \fi \nobreak
      \vrule height0.75em width0.5em depth0.25em\fi}

\newcommand{\abs}[1] {\left| {#1} \right|}

\newcommand{\prob} {\bbP}

\newcommand{\expect}[1] {\mathbb{E}\bracket{#1}}
\newcommand{\variance}[1] {\textrm{Var}\paren{#1}}

\newcommand{\paren}[1] {\left( {#1} \right)}
\newcommand{\bracket}[1] {\left[ {#1} \right]}
\newcommand{\curly}[1] {\left\{ {#1} \right\}}

\newcommand{\curlybig}[1] {\big\{ {#1} \big\}}

\newcommand{\bigO}{\mathcal{O}}

\newcommand{\card}[1] {\abs{#1}}

\def\whp/{whp}

\newcommand{\SetDef}[2] {\curlybig{\, {#1} \mid {#2} \,}}

\def\fref#1{Fig.~\ref{#1}}
\def\sref#1{Sec.~\ref{#1}}
\def\aref#1{Appendix~\ref{#1}}

\def\tabref#1{Table~\ref{#1}}
\def\asref#1{Assumption~\ref{#1}}

\usepackage{acronym}

\usepackage{siunitx}

\newcommand{\identitymatrix}{\bI}
\newcommand{\zeromatrix}{\bzero}

\newcommand{\intsupto}[1]{\{1,\dots,{#1}\}}

\newcommand{\AlgCmd}[1] {\textbf{#1}}

\newenvironment{alglevel}[1]%
  {\begin{list}{}%
          {\setlength{\leftmargin}{#1}}%
          \item[]%
  }
  {\end{list}}

\newcommand{\SketchIdx} {\calS}

\newcommand{\SketchIdxSnap}[1] {\SketchIdx({#1})}

\newcommand{\SketchIdxBirth} {\overline{\calS}}

\newcommand{\initialsketchsize}{N'}

\newcommand{\samplefunc}[2]{\textbf{Sample}(#1, #2)}

\newcommand{\inferfunc}[2]{\textbf{Infer}\paren{#1, #2}}

\newcommand{\mainalg}{\textbf{Main-Algorithm}}

\newcommand{\NBAlg}[1]{\textbf{Static-Cluster}(#1)}
\newcommand{\NBAlgnoparams}{\textbf{Static-Cluster}}

\newcommand{\growshrinkf}{f}
\newcommand{\NumNodes} {N}
\newcommand{\clustsizedef} {n}
\newcommand{\clustsketchsizedef} {n'}

\newcommand{\benchperiod} {\tau}
\newcommand{\benchphase} {\phi}

\newcommand{\NBEig}[1] {\lambda_{#1}}

\newcommand{\ClustGraph}[1] {G_{#1}}

\newcommand{\Clust}[1] {\widehat{C}_{#1}}
\newcommand{\ClustSet} {\widehat{\calC}}

\newcommand{\ClustSnapSet}[1] {\widehat{\calC}({#1})}
\newcommand{\ClustSnap}[2] {\widehat{C}_{#1}({#2})}

\newcommand{\ClustEstUnmatched}[1] {\overline{C}_{#1}}

\newcommand{\ClustEst}[1] {C_{#1}}
\newcommand{\ClustSetEst} {\calC}
\newcommand{\ClustSketchEst}[1] {C'_{#1}}
\newcommand{\ClustSketchSetEst} {\calC'}
\newcommand{\ClustSnapSketchSetEst}[1] {\calC'(#1)}
\newcommand{\ClustSnapSketchEst}[2] {C'_{#1}(#2)}
\newcommand{\ClustSnapSetEst}[1] {\calC({#1})}
\newcommand{\ClustSnapEst}[2] {C_{#1}({#2})}
\newcommand{\NumClustEst} {q}
\newcommand{\NumClustSnapEst}[1] {q(#1)}

\newcommand{\ClustIterEst} {C}
\newcommand{\ClustIterSketchEst} {C'}

\newcommand{\Modularity} {Q}

\newcommand{\norminv}[1]{\Phi^{-1}\!\paren{#1}}
\newcommand{\probofsuccess}{\alpha}

\newcommand{\snaptime}{t}
\newcommand{\snaptimeminusone}{{t\!-\!1}}
\newcommand{\period}{{T}}

\newcommand{\NumClust} {{\widehat{q}}}

\newcommand{\AlgCurrClustCount} {r}
\newcommand{\NumClustSnap}[1] {\widehat{q}(#1)}

\newcommand{\MaxNodes} {\overline{N}}
\newcommand{\MaxClust} {\overline{q}}

\newcommand{\SketchSize} {N'}
\newcommand{\SketchClustSize} {n'}

\newcommand{\NodeNewSnap}[1] {V^{+}({#1})}
\newcommand{\NodeDelete} {V^{-}}
\newcommand{\NodeDeleteSnap}[1] {V^{-}({#1})}
\newcommand{\NodeBirth} {V_{\textrm{birth}}}
\newcommand{\NodeBirthEst} {\hat{V}_{\textrm{birth}}}
\newcommand{\NodeBirthEstSnap}[1] {\NodeBirthEst({#1})}

\newcommand{\Nsamp} {{N'}}

\newcommand{\squarederrnormagree}{E_{\tilde{A}}}
\newcommand{\normagree} {\tilde{A}}
\newcommand{\normagreesnap}[1] {\tilde{A}(#1)}
\newcommand{\normagreedyn}[1]{\normagree^t}

\newcommand{\agree}[2] {\agreesymbol\paren{#1,#2}}
\newcommand{\agreesymbol} {A}

\newcommand{\birthdeathgap}{\gamma}

\newcommand{\movednodeset}[2]{\Node_{{#1} \rightarrow {#2}}}
\newcommand{\movednodesetsnap}[3]{\Node_{{#1} \rightarrow {#2}}({#3})}

\newcommand{\maxmovednodes}{\overline{m}}

\newcommand{\nummovednodessketch}[2]{m'_{{#1} \rightarrow {#2}}}

\newcommand{\nummovednodesintoclustsketch}[1]{\growshrinkmovednodes'_{(\cdot \rightarrow {#1})}}
\newcommand{\nummovednodesoutofclustsketch}[1]{\growshrinkmovednodes'_{({#1} \rightarrow \cdot)}}

\newcommand{\nminscaleresult}{\overline{n}_\textrm{min}}

\newcommand{\maxintraedges}{\widehat{m}}

\newcommand{\growshrinkmovednodes}{m}

\newcommand{\SplitSizeThresh} {a}
\newcommand{\DetectThresh} {d}

\newcommand{\AvgDeg} {c}

\newcommand{\clustsimdyn}[3] {s_{{#1},{#2}}(#3)}

\newcommand{\clustsimdynbinuin} {S^{\clustiter}_{in}}
\newcommand{\clustsimdynbinuout} {S^{\clustiter}_{out}}
\newcommand{\clustsimdynvar}[3] {\variance{\clustsimdyn{#1}{#2}{#3}}}

\newcommand{\clusteg} {u} 
\newcommand{\clustegtwo} {u'} 
\newcommand{\clustiter} {v} 
\newcommand{\clustitertwo} {w} 

\newcommand{\nodeeg} {i}
\newcommand{\nodeegtwo} {j}

\newcommand{\clustsizemin} {n_\textrm{min}}

\newcommand{\infersuffnormalmean} {\mu_{\Delta s}}
\newcommand{\infersuffnormalstd} {\sigma_{\Delta s}}
\newcommand{\infersuffnormalvar} {\infersuffnormalstd^2}
\newcommand{\infersuffnormalmeanbound} {\overline{\mu}}
\newcommand{\infersuffnormalvarbound} {\overline{x}}
\newcommand{\infersuffnormalboundrv} {f}

\newcommand{\pin} {p_{\textrm{in}}}
\newcommand{\pout} {p_{\textrm{out}}}

\newcommand{\birthprob} {\beta}
\newcommand{\probofsuccesstwo}{\beta'}

\newcommand{\pinest} {\widehat{p}_{\textrm{in}}}
\newcommand{\pineststddev} {\sigma_{\pinest}}
\newcommand{\pinestbin} {\widehat{P}_{\textrm{in}}}
\newcommand{\pinestupperbound}{p^+}
\newcommand{\pinestlowerbound}{p^-}
\newcommand{\pinestupperlowerbound}{p^{\pm}}

\newcommand{\pinterclustest}[3] {\widehat{p}_{{#1},{#2}}(#3)}
\newcommand{\pinterclustestnotime}[2] {\widehat{p}_{{#1},{#2}}}
\newcommand{\pinterclustestbin}[2] {\widehat{P}_{{#1},{#2}}}

\newcommand{\mergediffmean} {\mu^-}
\newcommand{\mergesummean} {\mu^+}
\newcommand{\mergesumdiffmean} {\mu^{\pm}}

\newcommand{\mergediffsumstd} {\sigma_{m}}

\newcommand{\pinterclust}[3] {p_{{#1},{#2}}(#3)}
\newcommand{\pinterclustnotime}[2] {p_{{#1},{#2}}}

\newcommand{\birthstdest}[1] {\widehat{\sigma}}

\newcommand{\Graph} {G}
\newcommand{\SubGraph} {G'}
\newcommand{\GraphSketch} {G'}
\newcommand{\GraphSnap}[1] {G({#1})}
\newcommand{\GraphSketchSnap}[1] {\GraphSketch({#1})}
\newcommand{\GraphBirthEst} {\overline{G}}

\newcommand{\GraphInducedBy}[2] {{#1}\!\bracket{#2}}

\newcommand{\Node} {V}
\newcommand{\NodeSnap}[1] {V({#1})}

\newcommand{\Edge} {E}
\newcommand{\EdgeSnap}[1] {{\Edge}({#1})}

\newcommand{\Adj} {\bA}

\newcommand{\mum} {m_{\textrm{um}}}
\newcommand{\mm} {m_{\textrm{m}}}

\newcommand{\DegMat} {\bD}

\newcommand{\EigVal}[1] {\lambda_{#1}}

\newcommand{\NBmatrixSurrogate}{\bB'}

\newacro{sbm}[sbm]{stochastic block model}
\newacro{SBM}[SBM]{Stochastic Block Model}
\newacro{DCSBM}[DCSBM]{Degree Corrected SBM}
\newacro{SPIN}[SPIN]{\textit{SamPling Inversely proportional to Node degree}}
\newacro{SVD}[SVD]{Singular Value Decomposition}
\newacro{ESPRA}[ESPRA]{\textit{Evolutionary clustering based on Structural Perturbation and Resource Allocation similarity}}

\def\StdInd/{(Independent)}
\def\Dinh/{\cite{5403845}}
\def\Yang/{\cite{Yang2011}}

\def\alglevelone{0.5cm}
\def\algleveltwo{1cm}
\def\alglevelthree{1.5cm}

\usepackage{enumitem}

\usepackage{graphicx}
\usepackage{dcolumn}
\usepackage{bm}

\begin{document}

\preprint{APS/123-QED}

\title{Sketch-based community detection in evolving networks}

\author{Andre Beckus}
\affiliation{%
Department of Electrical and Computer Engineering, University of Central Florida, Orlando, FL 32816 USA.
}%
 
\author{George K. Atia}%
\affiliation{%
Department of Electrical and Computer Engineering and Department of Computer Science, University of Central Florida, Orlando, FL 32816 USA.
}%

\date{\today}

\begin{abstract}

We consider an approach for community detection in time-varying networks. At its core, this approach maintains a small sketch graph to capture the essential community structure found in each snapshot of the full network. We demonstrate how the sketch can be used to explicitly identify six key community events which typically occur during network evolution: growth, shrinkage, merging, splitting, birth and death. Based on these detection techniques, we formulate a community detection algorithm which can process a network concurrently exhibiting all processes. One advantage afforded by the sketch-based algorithm is the efficient handling of large networks. Whereas detecting events in the full graph may be computationally expensive, the small size of the sketch allows changes to be quickly assessed. A second advantage occurs in networks containing clusters of disproportionate size. The sketch is constructed such that there is equal representation of each cluster, thus reducing the possibility that the small clusters are lost in the estimate. We present a new standardized benchmark based on the stochastic block model which models the addition and deletion of nodes, as well as the birth and death of communities. When coupled with existing benchmarks, this new benchmark provides a comprehensive suite of tests encompassing all six community events. We provide analysis and a set of numerical results demonstrating the advantages of our approach both in run time and in the handling of small clusters.

\end{abstract}

\maketitle

\section{Introduction}\label{sec:Intro}

The detection of community structure in networks has garnered a great deal of attention, leading to a vast array of algorithms.
Much of the focus has been on static networks, where the goal is to identify groups of nodes within which connections are dense and between which connections are relatively sparse.
However, it is often the case that networks evolve with time.
For example, edges in social media networks appear and disappear to reflect ever-changing friendships, and gene expression networks continuously evolve in response to external stimuli
\cite{Aggarwal:2014:ENA:2620784.2601412,5562773}.
In this dynamic setting, new sequential algorithms are needed to track the community structure underlying each temporal snapshot of the network.
Here, we propose a sketch-based approach.

Sketching involves the construction of a small synopsis of a full dataset \cite{Cormode:11}.
Notably, this technique has been used in static community detection \cite{8955821,Beckus_MLSP:19}, where a sketch sub-graph is generated by sampling nodes from the full network.
The sketch is clustered using an existing community detection algorithm, and the community membership of the nodes in the full network are inferred based on the estimated communities in the sketch.
Here, we propose the use of an evolving sketch to detect and handle the six canonical community events observed in dynamic networks \cite{SHANG201670}: growth, shrinkage, merging, splitting, birth and death.
This dynamic approach addresses two pervasive issues in community detection.

One important concern in community detection is the ability to process large graphs.
Many static methods become infeasibly slow when processing a large network, thus motivating a search for efficient algorithms \cite{PhysRevE.70.066111}.
The extra time dimension inherent to the dynamic setting only makes this search for efficiency more pressing.
However, time-evolving networks also offer a distinct advantage not found in the static domain.
Specifically, evolving networks often possess temporal smoothness in which the community structure changes gradually \cite{DAKICHE20191084}.
In this case, previous snapshots offer prior information which can aid in the clustering of subsequent snapshots.
We present a method which relies on a small sketch to convey information regarding previous snapshots.
By using a small sketch, the algorithm can detect the main community events without requiring the full graph to be examined, thus reducing the required computational complexity.
If the sketch size and number of clusters are fixed, the complexity of our algorithm scales linearly in network size.

Another typical issue found in community detection is the detection of small clusters \cite{zhang2012community}.
If a community shrinks too small, it may become lost, i.e., the community may be absorbed into a larger community in the estimated partition.
We show that once a community is captured in the sketch, it can be tracked even if the community becomes very small.

We use dynamic benchmarks as a means for evaluating the proposed algorithm with respect to the canonical network events.
The first four events are included in the benchmarks of \cite{PhysRevE.92.012805},
which are based on the well-known \ac{SBM} \cite{HOLLAND1983109}.
Here, we propose a new dynamic \ac{SBM} benchmark which captures the last two events of birth and death.
An important feature of this proposed benchmark is that the size of the network varies with time, a characteristic not found in the existing benchmarks.
In addition to modeling the birth event, this benchmark incrementally adds new nodes to the network which join existing communities, a feature also not seen in \cite{PhysRevE.92.012805}.

This paper is organized as follows.
In \sref{sec:RelatedWork}, we summarize existing community detection algorithms for evolving networks.
\sref{sec:DynGraphs} describes the network model, and
\sref{sec:evolutionprocesses} summarizes \ac{SBM} benchmarks which capture key evolutionary processes.
In \sref{sec:SketchEvents}, we describe the sketch-based approach, and formulate techniques by which sketches can detect events and track evolutionary processes.
\sref{sec:SummaryAlg} presents the proposed algorithm based on these tracking techniques.
We analyze the algorithm in \sref{sec:analysis}, present numerical results in \sref{sec:results}, and conclude in \sref{sec:conclusion}.
\aref{sec:staticclust} describes the static clustering used as a part of the main algorithm, and
\aref{sec:mainalgdetails} provides details on the main algorithm itself.
\aref{sec:proofs} derives the results found in the analysis of \sref{sec:analysis}.
\aref{sec:comparisonalgorithms} provides additional details regarding the algorithms we compare against in the numerical results.

\section{Related work: community detection in evolving networks} \label{sec:RelatedWork}

A number of algorithms have been proposed for community detection in evolving networks (see \cite{DAKICHE20191084,Rossetti:2018:CDD:3186333.3172867} for comprehensive surveys).
One straightforward approach entails the independent clustering of each snapshot using a static clustering algorithm.
The communities in the current snapshot are matched to the previous communities such that there is continuity in the community identities.
This category of algorithm contains a number of variants beginning with the classic work of \cite{Hopcroft5249}.

More recently, many algorithms take a more sophisticated ``dependent'' approach, in which previous snapshots are accounted for in the clustering of the current snapshot.
These algorithms have the potential to outperform independent community detection algorithms, since they incorporate previous knowledge \emph{directly} in the clustering step.

One approach commonly seen in this category is the representation of each snapshot using a compact graph.
In \cite{5403845}, a small weighted graph is constructed after clustering a given snapshot, with each community represented by a single ``supernode''.
The weights of the edges between supernodes indicate the cumulative number of edges between the corresponding communities.
These supernodes are then incorporated into the next snapshot's graph, thus carrying forward information from the previous estimates.
A similar idea can be seen in dynamic methods built around the static Louvain algorithm \cite{1742-5468-2008-10-P10008}, for example as seen in \cite{HE201587}.
The extension of the Louvain algorithm to time-varying networks follows naturally from its reliance on supernodes.
Our approach also uses a small representative graph, however using an altogether different idea of sketching, as described in \sref{sec:SketchEvents}.

The model used in this paper is based on the \ac{SBM} \cite{HOLLAND1983109}.
Several recent algorithms have been developed based on dynamic \ac{SBM}-based models.
The dynamic models of \cite{Yang2011,PhysRevX.6.031005} specify that nodes move between a fixed set of communities according to a stationary transition probability matrix.
In addition to allowing the movement of nodes between communities, the models of \cite{6758385,penskyDSBM2019} also allow the edge probabilities of the communities to vary.
Nonetheless, these works focus on the case where individual nodes only change community membership, i.e., the communities undergo the grow and shrink processes.
Although \cite{9721420} is able to track communities which are also merging and splitting, it still does not allow varying numbers of nodes across the snapshots.
The algorithm of \cite{matias2017statistical} allows nodes to join or leave the graph, but requires that all snapshots be known when invoking the algorithm.
We emphasize that our proposed algorithm is online in nature, i.e., it performs community detection iteratively on one snapshot at a time, while carrying forward the clustering results from previous snapshots.

\section{Temporal Network Model} \label{sec:DynGraphs}

At time $\snaptime$, the network snapshot is represented by graph $\GraphSnap{\snaptime} \!=\! \paren{\NodeSnap{\snaptime},\EdgeSnap{\snaptime}}$, where $\NodeSnap{\snaptime}$ is the set of nodes in existence at time $\snaptime$, and $\EdgeSnap{\snaptime}$ is the set of edges between these nodes.
Let $\ClustSnapSet{\snaptime} = \SetDef{\ClustSnap{\clusteg}{\snaptime}}{\clusteg \in \intsupto{\NumClustSnap{\snaptime}}}$ be the partition at time $\snaptime$, with $\ClustSnap{\clusteg}{\snaptime}$ denoting the set of nodes in community $\clusteg$, and $\NumClustSnap{\snaptime}$ the number of communities.

In each snapshot, an edge exists between nodes within a community with probability $\pin$.
Nodes in community $\clusteg$ are connected to nodes in a different community $\clustegtwo$ with probability $\pinterclust{\clusteg}{\clustegtwo}{\snaptime}$.
An evolutionary process may vary the intercommunity edge density $\pinterclust{\clusteg}{\clustegtwo}{\snaptime}$ so long as the resulting graph adheres to the \ac{SBM}. We discuss one such process in \sref{sec:mergeshrinkprocess}.
A pair of communities $\clusteg, \clustegtwo$ are considered to be merged if
\begin{align} \label{eqn:mergethresh}
\pin - \pinterclust{\clusteg}{\clustegtwo}{\snaptime} <
\sqrt{
\frac{2 \paren{\pin + \pinterclust{\clusteg}{\clustegtwo}{\snaptime}}}
{\card{\ClustSnap{\clusteg}{\snaptime}} + \card{\ClustSnap{\clustegtwo}{\snaptime}}}}.
\end{align}
When communities are of equal size, this condition corresponds to the asymptotic weak detectability limit (see \cite{PhysRevE.92.012805} for a discussion of this bound in the context of merging and splitting communities).
For simplicity, here, we average the community sizes when they are of unequal size.

The following events may occur at time $\snaptime$.
\begin{itemize}
    \item \textbf{Node movement between communities}
          A set of nodes $\movednodesetsnap{\clusteg}{\clustegtwo}{\snaptime}$ belonging to community $\clusteg$ may move to community $\clustegtwo$. The edges connected to these nodes are regenerated according to the \ac{SBM} based on the new community memberships.
    \item \textbf{New nodes and community birth}
          A set of nodes $\NodeNewSnap{\snaptime}$ may join the graph.
          A subset $\NodeBirth(\snaptime) \subseteq \NodeNewSnap{\snaptime}$ of these nodes join new communities.
          The remaining nodes $\NodeNewSnap{\snaptime} \setminus \NodeBirth(\snaptime)$ join existing communities.
          The edges of nodes in $\NodeNewSnap{\snaptime}$ are generated according to the \ac{SBM}.
    \item \textbf{Removed nodes and community death}
          A set of nodes $\NodeDeleteSnap{\snaptime}$ may be removed from the graph. Death occurs when all nodes in a particular community are removed.
    \item \textbf{Merge and split of communities}
          The merge event occurs for communities $\clusteg, \clustegtwo$ when $\pinterclust{\clusteg}{\clustegtwo}{\snaptime}$ increases such that \eqref{eqn:mergethresh} becomes true.
          Likewise, the split event occurs when $\pinterclust{\clusteg}{\clustegtwo}{\snaptime}$ decreases such that \eqref{eqn:mergethresh} becomes false.
\end{itemize}

Note that many of the model variables are functions of time $\snaptime$.
Where there is no ambiguity, we omit this time parameter to simplify the exposition.

\section{Evolutionary Processes: Benchmarks}
\label{sec:evolutionprocesses}
For the purpose of illustrating and analyzing the proposed algorithm, we consider here specific examples of evolutionary processes.
These are realized by four benchmark networks, i.e., parameterized sequences of snapshots with known community partitions for validating and comparing community detection algorithms.
The grow-shrink and merge-split benchmarks are defined in \cite{PhysRevE.92.012805}, whereas we present the birth-death process here for the first time.

Each benchmark consists of an evolving network containing $2n$ total nodes.
The underlying process is driven by a periodic triangular waveform
\begin{align}
x(t) &=
\begin{cases}
2 t^{*}, & 0 \leq t^{*} < 1 / 2, \\
2-2 t^{*}, & 1 / 2 \leq t^{*} < 1,
\end{cases}
\end{align}
where
\begin{align}
t^{*} &\equiv(t / \tau + \phi) \bmod 1,
\end{align}
$\tau$ is the period of the waveform, and $\phi$ controls the phase of the waveform.
We will assume that $\phi\!=\!0$ unless otherwise specified.

\subsection{Grow-shrink benchmark} \label{sec:growshrinkprocess}

The grow-shrink benchmark moves nodes between a pair of communities denoted $A$ and $B$, thus growing and shrinking the communities.
At each time step the first community contains
\begin{align}
n_A = n - nf[2x(t + \benchperiod/4)-1]
\end{align}
nodes, whereas the second community contains $n_B\!=\!2n\!-\!n_A$ nodes.
Nodes lost from the first community are transferred to the second community, and vice-versa.
The parameter $\growshrinkf \in [0,1]$ controls the variation in community sizes.
For $\snaptime \in \{0, \benchperiod/2, \benchperiod\}$ the sizes of the communities are equal.
At time $\snaptime=\benchperiod/4$, a fraction $\growshrinkf$ of nodes in community $A$ will have moved to community $B$, whereas at time $\snaptime\!=\!3 \benchperiod / 4$ the opposite holds.

\subsection{Merge-split benchmark} \label{sec:mergeshrinkprocess}

The merge-split benchmark has two communities denoted $A$ and $B$, each of size $\clustsizedef$, with intracommunity edge density $\pin$.
Initially, the intercommunity edge density is $\pinterclust{A}{B}{0}=\pout$.
New edges are gradually added between the two communities until they are completely merged at time $\snaptime\!=\!\benchperiod/2$ with $\pinterclust{A}{B}{\benchperiod/2}=\pin$.
Then, the process reverses and the new edges are removed until the communities are completely split again at time $\snaptime\!=\!\benchperiod$.

The intercommunity edges are placed in the following way.
The number of intercommunity edges $\mum$ in the unmerged state are drawn according to a binomial distribution with parameters $n^2$ and $\pout$.
The number of edges $\mm$ in the merged state is similarly drawn, except using probability $\pin$.
The number of edges at time $\snaptime$ is then determined by
\begin{align}
m^*(\snaptime) = [1-x(\snaptime)]\mum + x(\snaptime)\mm,
\end{align}
where the edges are placed uniformly at random.
In this way, the edge density between the two communities is $\pinterclust{A}{B}{\snaptime} = m^*(\snaptime) / \clustsizedef^2$.
The communities are considered merged at the detectability limit \eqref{eqn:mergethresh}.

\subsection{Birth-death benchmark}

\begin{figure}
\centering
\includegraphics[scale=1]{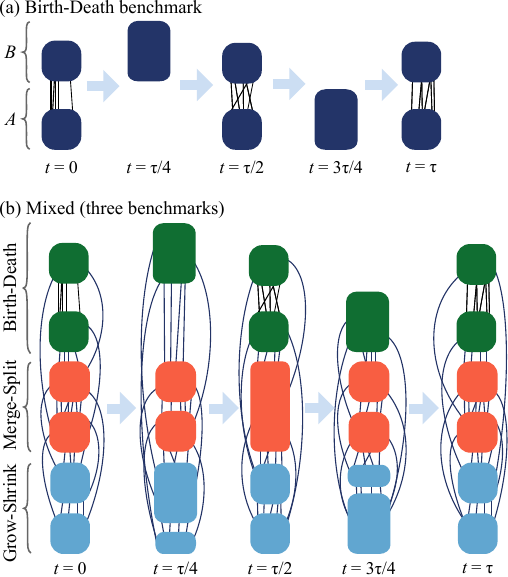}
\caption{
(a) Schematic representation of the birth-death benchmark, showing the two communities labeled $A$ and $B$.
(b) Schematic representation of the mixed benchmark, which stacks the grow-shrink, merge-split, and birth-death benchmarks.
}
\label{fig:BenchSchematic}
\end{figure}

We now propose a new benchmark which realizes the birth and death of communities, as well as the addition and removal of nodes from the network.
A schematic diagram of the birth-death benchmark is shown in \fref{fig:BenchSchematic}(a).
The benchmark contains two communities which pass into and out of existence.
The size of the first community is
\begin{align}
n_A &= 
\begin{cases}
0,                  & x(t+\benchperiod/4) \geq 1-\birthdeathgap/2, \\
n \bracket{1-x(t+\benchperiod/4)}, & otherwise,
\end{cases}
\end{align}
where $n_A\!=\!0$ designates a non-existent community, and parameter $\birthdeathgap \in [0,1]$ controls the minimum size of the community.
The community starts at time $\snaptime\!=\!0$ with $n/2$ nodes.
Nodes are removed from the network, until the community shrinks to size $\birthdeathgap n / 2$ at time $\snaptime\!=\!\benchperiod(1-\birthdeathgap) / 4$.
At this point, the community dies and all of its remaining nodes are deleted from the network.
At time $\snaptime=\benchperiod(1+\birthdeathgap) / 4$, a new set of $\birthdeathgap n / 2$ nodes is added to the network and used to re-create the community.
New nodes are gradually created and added to the community until it reaches size $\clustsizedef$.
At this point, nodes are again removed from the community until it contains $n/2$ nodes, and the process repeats.

The second community is of size
\begin{align}
n_B &= 
\begin{cases}
0,         & x(t+\benchperiod/4) < \birthdeathgap/2 \\
n \, x(t+\benchperiod/4), & otherwise.
\end{cases}
\end{align}
This community undergoes essentially the same process as the first community except with a phase shift of $\benchperiod/2$.

\subsection{Mixed benchmark}
\label{sec:mixedbenchmarks}

\begin{figure}
\centering
\includegraphics[scale=1]{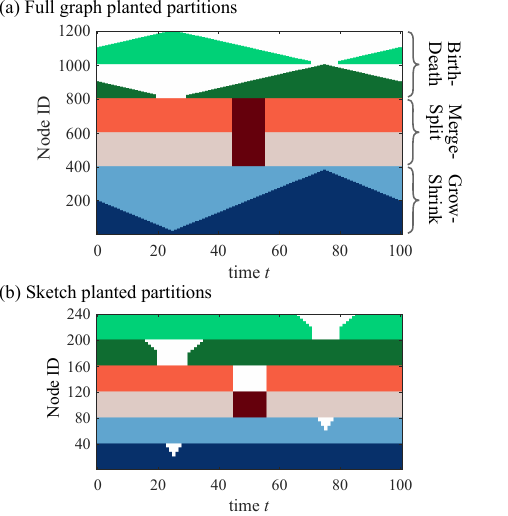}
\caption{
(a) Planted partitions for full graphs of the mixed benchmark network. Each vertical slice indicates the planted partition at time $\snaptime$.
The model parameters are $\clustsizedef\!=\!200,\NumClust\!=\!6,\growshrinkf\!=\!0.9, \birthdeathgap\!=\!0.2$
(b) Sketches produced with $\clustsketchsizedef\!=\!40$.
Each vertical slice indicates the planted partitions in sketch $\SketchIdxSnap{\snaptime}$.
White regions indicate that the corresponding node does not exist at time $\snaptime$.}
\label{fig:birthdeathdiagram}
\end{figure}

To model concurrent processes capturing all of the events,
we present a mixed benchmark which is created by ``stacking'' the grow-shrink, merge-split, and birth-death benchmarks.
A schematic of this mixed benchmark is shown in \fref{fig:BenchSchematic}(b).
The benchmark has a maximum of $6\clustsizedef$ nodes.
The first $4\clustsizedef$ nodes contain the grow-shrink and merge-split benchmarks as previously described, whereas the last $2\clustsizedef$ nodes participate in the birth-death process (the actual number of nodes varies with time due to addition and deletion of nodes in the birth-death benchmark).
We show an example of this mixed benchmark in \fref{fig:birthdeathdiagram}(a).

\section{Sketch-based Tracking of Evolutionary Processes} \label{sec:SketchEvents}

Our algorithm relies on a small representative sketch of the full network.
The sketch captures important information which can be used to detect network events and track the processes by which the network evolves.
Meanwhile, the smaller size of the sketch allows these checks to be performed quickly without requiring a complete assessment of the entire network.
The estimates of the communities in snapshot $\snaptime$ are $\ClustSnapSetEst{\snaptime} = \SetDef{\ClustSnapEst{\clusteg}{\snaptime}}{\clusteg \in \intsupto{\NumClustSnapEst{\snaptime}}}$, where $\NumClustSnapEst{\snaptime}$ is the estimated number of communities at time $\snaptime$.
We first describe the sketch, and then describe how this sketch can be used to detect specific events.

The sketch consists of a set of nodes sampled from the full network.
At each time step, this set is updated such that it contains an equal number of nodes $\clustsketchsizedef$ from each community.
The set of nodes in the sketch at time $\snaptime$ is denoted $\SketchIdxSnap{\snaptime}$, and the subset of these nodes from community $\clusteg$ is denoted $\ClustSnapSketchEst{\clusteg}{\snaptime} = \SketchIdxSnap{\snaptime} \cap \ClustSnapEst{\clusteg}{\snaptime}$.
We refer to this as sketch community $\clusteg$.
An example sketch time series is shown in \fref{fig:birthdeathdiagram}(b), where nodes have been sampled from the mixed benchmark shown in \fref{fig:birthdeathdiagram}(a).

For this example, we build the sketches using knowledge of the planted community partitions.
The proposed algorithm has no such knowledge, and therefore must build the sketches based on estimates of the true communities.
We will present an actual sketch produced by the proposed algorithm in \sref{sec:resultssimultaneous}.

\subsection{Inferring community membership of nodes}
\label{sec:infer}
We show in this section how the sketch may be used to infer community membership of any node $\nodeeg \!\in\! \NodeSnap{\snaptime}$ in network snapshot $\GraphSnap{\snaptime}$.
To this end, we calculate
\begin{align} \label{eqn:simmetric}
\clustsimdyn{\nodeeg}{\clustiter}{\snaptime}
= \frac{ \card{ \SetDef{(\nodeeg,\nodeegtwo) \in \EdgeSnap{\snaptime}}{\nodeegtwo \in \ClustSnapSketchEst{\clustiter}{\snaptimeminusone}} } }{\card{ \ClustSnapSketchEst{\clustiter}{\snaptimeminusone} } }
\end{align}
to evaluate the connectivity of node $\nodeeg$ to each sketch community $\clustiter$.
Let $\clusteg$ be the true community assignment of node $\nodeeg$.
Since
\begin{align} \label{eqn:expectsimmetric}
\expect{ \clustsimdyn{\nodeeg}{\clustiter}{\snaptime} }
=
\begin{cases}
\pin,  &\clustiter = \clusteg, \\
\pinterclust{\clusteg}{\clustiter}{\snaptime}, &\clustiter \ne \clusteg,
\end{cases}
\end{align}
it follows that $\clustsimdyn{\nodeeg}{\clustiter}{\snaptime}$ provides a point estimate of the probability that there is an edge between node $\nodeeg$ and any node $\nodeegtwo \in \ClustSnapSketchEst{\clustiter}{\snaptimeminusone}$.
Node $\nodeeg$ can then be assigned to the community $\clustegtwo$ with which connectivity is greatest, i.e., where
\begin{align} \label{eqn:estclust}
\clustegtwo = \argmax_{1 \le \clustiter \le \NumClustSnapEst{\snaptime}} \clustsimdyn{\nodeeg}{\clustiter}{\snaptime}.
\end{align}
The proposed algorithm uses \eqref{eqn:estclust} to assign communities to new nodes joining the network, as well as to identify nodes which have changed community membership.

We finish this section by noting that the variance in $\clustsimdyn{\nodeeg}{\clustiter}{\snaptime}$ is
\begin{align} \label{eqn:varsimmetric}
\clustsimdynvar{\nodeeg}{\clustiter}{\snaptime}
=
\begin{cases}
\frac{\pin(1-\pin)}{ \card{ \ClustSnapSketchEst{\clustiter}{\snaptimeminusone} } },  &\clustiter = \clusteg, \\
\frac{\pinterclust{\clusteg}{\clustiter}{\snaptime}[1-\pinterclust{\clusteg}{\clustiter}{\snaptime}]}{\card{ \ClustSnapSketchEst{\clustiter}{\snaptimeminusone} }},  &\clustiter \ne \clusteg.
\end{cases}
\end{align}
The variance grows as the sketch communities shrink, thus motivating the use of equal-sized communities in the sketch.

\subsection{Detecting the split event} \label{sec:splitprocess}

Suppose that community $\clusteg$ is undergoing a split into two separate communities $\clusteg$ and $\clustegtwo$.
To detect the emerging clusters we can use the spectrum of the non-backtracking matrix as described in \cite{Krzakala20935}.
Let $\Graph_{\clusteg}$ be the sub-graph of $\GraphSnap{\snaptime}$ induced by the latest estimate $\ClustSnapEst{\clusteg}{\snaptimeminusone}$, and $\Adj$ be the adjacency matrix of $\Graph_{\clusteg}$.
Given diagonal matrix $\DegMat$ containing the degrees of nodes in $\Adj$, and identity matrix $\identitymatrix$, define
\begin{align} \label{eqn:NBsurrogate}
    \NBmatrixSurrogate =
    \begin{pmatrix}
    \zeromatrix & \DegMat\!-\!\identitymatrix \\
    -\identitymatrix & \Adj
    \end{pmatrix}.
\end{align}

Suppose the emerging communities are each of size $\clustsizedef$, and define
$\NBEig{1}, \NBEig{2}$ as the largest and second largest eigenvalues of $\NBmatrixSurrogate$, respectively.
If
\begin{align} \label{eqn:detectthresh}
\clustsizedef \paren{\pin - \pinterclust{\clusteg}{\clustegtwo}{\snaptime}}^2 >
\pin + \pinterclust{\clusteg}{\clustegtwo}{\snaptime},
\end{align}
then in the limit as $\clustsizedef \!\rightarrow\! \infty$ with $\clustsizedef \pin$ and $\clustsizedef \pinterclust{\clusteg}{\clustegtwo}{\snaptime}$ constant,
$\NBEig{1} \!\rightarrow\! \clustsizedef \paren{ \pin+\pinterclust{\clusteg}{\clustegtwo}{\snaptime}}$
and
$\lambda_2 \!\rightarrow\! \clustsizedef \paren{ \pin-\pinterclust{\clusteg}{\clustegtwo}{\snaptime}}$
such that
\cite{Krzakala20935}
\begin{align} \label{eqn:splitdecision}
\NBEig{2} > \sqrt{\NBEig{1}}.
\end{align}

\begin{figure}
\centering
\includegraphics[scale=1]{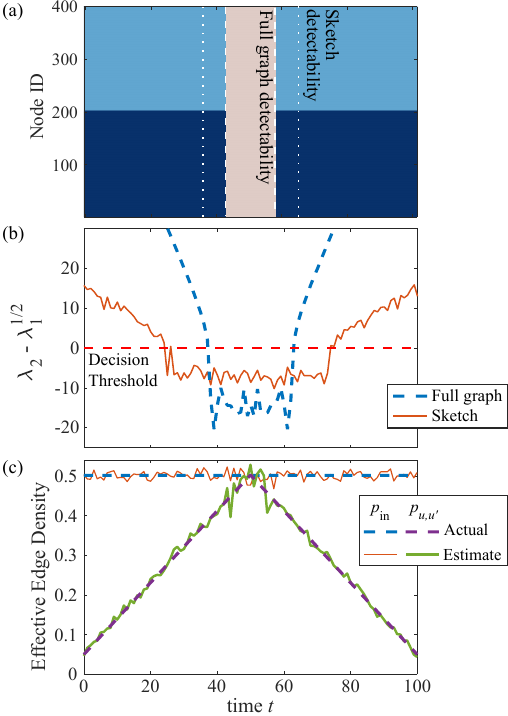}
\caption{
(a) Planted partitions of the merge-split benchmark.
For reference, the detectability limits that formally define the splits in the benchmark are shown as vertical dashed and dotted lines.
Network parameters are $\NumClust\!=\!2, \clustsizedef\!=\!200,\pin\!=\!0.5,\pout\!=\!0.05$.
(b) Actual and estimated values of $\NBEig{2} \!-\! \sqrt{\NBEig{1}}$ at each time step. When the gap is greater than the decision threshold (horizontal dashed red line), the community is considered split.
The sketch is constructed using $\clustsketchsizedef\!=\!50$ nodes sampled uniformly at random from each of the two communities at each time step.
(c) Actual and estimated values of $\pin,\pinterclustnotime{\clusteg}{\clustegtwo}$ at each time step.
}
\label{fig:MergeSplitConcepts}
\end{figure}

Although condition \eqref{eqn:splitdecision} is only valid in the limit of infinite sized graphs, it can still serve as a reliable split indicator for a given sequence of network realizations.
We show an example of this in \fref{fig:MergeSplitConcepts}.
The planted partitions are shown in \fref{fig:MergeSplitConcepts}(a), and the dashed blue line in \fref{fig:MergeSplitConcepts}(b) shows the corresponding gap $\NBEig{2} \!-\! \sqrt{\NBEig{1}}$ for each time step.
The value of this gap increases as the process moves in either direction away from the fully merged state at $\snaptime\!=\!50$, and towards the fully split states at $\snaptime \in \{0,\benchperiod\}$.
Decision threshold \eqref{eqn:splitdecision} is shown as a horizontal dashed line.
As can be seen, the split is detected fairly close to the full graph detectability limit.

Rather than calculating the eigenvalues for the full network (at great computational cost), we propose to instead detect the split using the sketch.
We apply the same procedure as described above, but instead substitute $\ClustSnapEst{\clusteg}{\snaptimeminusone}$ with $\ClustSnapSketchEst{\clusteg}{\snaptimeminusone}$.
The estimate based on the sketch is shown in \fref{fig:MergeSplitConcepts}(b) as a solid orange line.
Note that the time of detection in the sketch diverges from that in the full graph as the community sizes in the sketch decrease.
We analyze this dependence on sketch size in \sref{sec:analysis}. 

\subsection{Detecting the merge event} \label{sec:mergeprocess}

Suppose that communities $\clusteg,\clustegtwo$ are merging.
To detect the merge event, we exploit the fact that the two communities are already known at time $\snaptimeminusone$.
This means that we can estimate $\pin$ and $\pinterclust{\clusteg}{\clustegtwo}{\snaptime}$ and use these estimates to directly check condition \eqref{eqn:mergethresh} to detect a merge event.
The sketch allows us to quickly calculate point estimates of the edge probabilities using the expressions
\begin{align}
\pinest
&= \frac{ 2 \sum_{\clustiter = 1}^{\NumClustSnapEst{\snaptime}} \card{ \SetDef{(i,j) \in \EdgeSnap{\snaptime}}{\nodeeg, \nodeegtwo \in \ClustSnapSketchEst{\clustiter}{\snaptimeminusone}} }}
{\sum_{\clustiter = 1}^{\NumClustSnapEst{\snaptime}} \card{\ClustSnapSketchEst{\clustiter}{\snaptimeminusone}} \paren{\card{\ClustSnapSketchEst{\clustiter}{\snaptimeminusone}}-1}}, \label{eqn:pinest}
\\
\pinterclustest{\clusteg}{\clustegtwo}{\snaptime}
&= \frac{\card{ \SetDef{(\nodeeg,\nodeegtwo) \!\in\! \EdgeSnap{\snaptime}}{\nodeeg \!\in\! \ClustSnapSketchEst{\clusteg}{\snaptimeminusone}, \nodeegtwo \in \ClustSnapSketchEst{\clustegtwo}{\snaptimeminusone}} }}
{\card{\ClustSnapSketchEst{\clusteg}{\snaptimeminusone}} \card{\ClustSnapSketchEst{\clustegtwo}{\snaptimeminusone}}}.
\end{align}

Figure~\ref{fig:MergeSplitConcepts}(c) shows the actual (dashed blue line) and estimated (solid orange line) values of $\pin$ for the example in \fref{fig:MergeSplitConcepts}(a).
The actual (dashed purple line) and estimated (solid green line) values of $\pinterclust{\clusteg}{\clustegtwo}{\snaptime}$ are also shown in the sample plot.
In both cases, the estimates track the actual values well.

\subsection{Detecting the birth event}

Consider a node $\nodeeg \in \NodeNewSnap{\snaptime}$, which is joining the network.
If the node joins an existing community $\clusteg$, then $\expect{ \clustsimdyn{\nodeeg}{\clusteg}{\snaptime} }=\pin$, and we can detect this occurrence by checking if $
\clustsimdyn{\nodeeg}{\clusteg}{\snaptime} 
\ge
\pinest - 3 \birthstdest{\clustiter},
$
where
$\pinest$ is the estimate from \eqref{eqn:pinest}, and
$
\birthstdest{\clusteg}
= \sqrt{ \pinest (1-\pinest) / \clustsketchsizedef }
$
is an estimate of the standard deviation of $\clustsimdyn{\nodeeg}{\clusteg}{\snaptime}$.
On the other hand, if $\nodeeg \in \NodeBirth(\snaptime)$, then the expectation $\expect{ \clustsimdyn{\nodeeg}{\clustiter}{\snaptime} }$ will equal the intercommunity edge density between the new community and \emph{any} existing community $\clustiter$.
This suggests that we can identify the set of nodes that are joining newborn communities using the expression
\begin{align} \label{eqn:birthnodenodeest}
\NodeBirthEstSnap{\snaptime} \!=\!
\left\{
i \in \NodeNewSnap{\snaptime}
\mid
\clustsimdyn{\nodeeg}{\clustiter}{\snaptime} 
<
\pinest - 3 \birthstdest{\clustiter},
\right.
\nn\\
\left.
\forall \clustiter \in \intsupto{\NumClustSnapEst{\snaptime}}
\right\}.
\end{align}

\section{Proposed Algorithm}
\label{sec:SummaryAlg}

We first discuss preliminaries.
The proposed algorithm invokes a function $\NBAlg{\Graph}$, which performs clustering of a static graph $\Graph$ to produce community estimates $\ClustSetEst = \curly{ \ClustEst{1}, \dots, \ClustEst{q} }$.
We implement this function using spectral techniques based on the non-backtracking matrix \cite{Krzakala20935}, along with enhancements to provide more robust estimation of the number of communities (details of the function are given in \aref{sec:staticclust}).
The computational cost of this function is dominated by the eigendecomposition, which is cubic in the size of graph $\Graph$.
We now summarize the main steps of the proposed algorithm.
We provide an assessment of the computational complexity for each step, and comment on the overall complexity at the end.
A detailed algorithm listing is provided in \aref{sec:mainalgdetails}.

\bigskip

\noindent$\mainalg$

\noindent \textit{Input:}
Initial sketch size $\initialsketchsize$.
Sketch community size $\clustsketchsizedef$.
Graph snapshots $\GraphSnap{\snaptime}, \snaptime=0,1,\dots$

\begin{enumerate}[topsep=0pt,noitemsep,label={(\arabic*)},ref=\theenumi]

    \item \textbf{Cluster initial snapshot.}
    Build sketch $\GraphSketch$ by sampling $\initialsketchsize$ nodes from $\GraphSnap{0}$ uniformly at random.
    Invoke $\NBAlg{\GraphSketch}$ to obtain community estimates $\ClustSketchSetEst$ for the sketch.
    Use \eqref{eqn:estclust} to infer the community memberships $\ClustSnapSetEst{0}$ of all nodes in $\GraphSnap{0}$ based on community estimates $\ClustSketchSetEst$.
    
    \textit{Complexity:}
    By executing $\NBAlgnoparams$ solely on the sketch, we reduce the running time of this expensive step to only
    $\bigO \paren{ \initialsketchsize^3 \!+\! \initialsketchsize \card{\NodeSnap{0}} }$.
    The first term corresponds to clustering of the sketch, and the second term corresponds to inference on the full graph. 
    \label{step:initclust}

    \item \AlgCmd{For} each snapshot $\GraphSnap{\snaptime}, \snaptime=1,2,\dots$ \AlgCmd{do}

        \item
        \begin{alglevel}{\alglevelone}
        \textbf{Update sketch.} Update the sketch to include $\clustsketchsizedef$ nodes sampled uniformly at random from each community.
        \end{alglevel}
        \label{step:loopstart}
        
        \item
        \begin{alglevel}{\alglevelone}
            \textbf{Birth detection.}
            Identify newborn communities by calculating $\NodeBirthEst$ as in \eqref{eqn:birthnodenodeest}.
            Since there may be more than one community born at the same time, we cluster the graph induced by $\NodeBirthEst$ using $\NBAlgnoparams$.
            To keep running time low, we use the same sketch-based approach as in Step~\ref{step:initclust}.

            \textit{Complexity:}
            The clustering of the nodes in $\NodeBirthEst$ incurs the dominant cost.
            We use a sketch consisting of $\clustsketchsizedef$ nodes from $\NodeBirthEst$, and so the clustering will take time $\bigO \paren{ \NumClustSnapEst{\snaptime}^3 \clustsketchsizedef^3 \!+\! \NumClustSnapEst{\snaptime} \clustsketchsizedef \card{\NodeSnap{\snaptime}} }$
        \end{alglevel}
        \label{step:birthdetect}

        \item
        \begin{alglevel}{\alglevelone}
            \textbf{Infer community membership of new and moved nodes.}
            Use the estimator \eqref{eqn:estclust} to infer community membership of each node $\nodeeg \in \NodeSnap{\snaptime} \setminus \NodeBirthEst$.
            Note that this set includes existing nodes, which may have changed community membership, as well as new nodes $\NodeNewSnap{\snaptime}$ which are joining existing communities.
            
            \textit{Complexity:}
            Calculation of the similarity metric $\clustsimdyn{\nodeeg}{\clusteg}{\snaptime}$ for a single community $\clusteg$ and single node $\nodeeg$ takes time $\bigO \paren{ \clustsketchsizedef }$.
            Therefore, this step is $\bigO \paren{ \NumClustSnapEst{\snaptime} \, \clustsketchsizedef \card{\NodeSnap{\snaptime}}}$ in total.
        \end{alglevel}
        \label{step:nodeinfer}
        
        \item
        \begin{alglevel}{\alglevelone}
            \textbf{Split detection.}
            For each community $\clusteg$, build graph $\GraphSketch$ induced by $\ClustSnapSketchEst{\clusteg}{\snaptime}$.
            From this induced graph, build $\NBmatrixSurrogate$ as defined in \eqref{eqn:NBsurrogate}.
            Calculate the eigenvalues $\NBEig{1},\NBEig{2}$ of $\NBmatrixSurrogate$.
            If $\NBEig{2} \!>\! \sqrt{\NBEig{1}}$, then a split event is declared.
            In this case, invoke $\NBAlg{\GraphSketch}$ to identify the emerging communities in the sketch, and then use \eqref{eqn:estclust} to identify nodes in the full graph belonging to these emerging communities.
            
            \textit{Complexity:}
            In the worst case, for each sketch community we must perform an eigendecomposition, estimate the partitions, and infer community membership in the full graph.
            Thus, this step is $\bigO \paren{\NumClustSnapEst{\snaptime} \clustsketchsizedef ( \clustsketchsizedef^2 + \card{\NodeSnap{\snaptime}})}$ in total.
        \end{alglevel}
        \label{step:splitdetect}
        
        \item
        \begin{alglevel}{\alglevelone}
            \textbf{Merge detection.}
            For each pair of communities $\clusteg, \clustegtwo$,
            consider the communities merged if
            \begin{align} \label{eqn:algdetectmerge}
                \pinest - \pinterclustestnotime{\clusteg}{\clustegtwo}
                < \DetectThresh
                \sqrt{\frac{2 \paren{ \pinest + \pinterclustestnotime{\clusteg}{\clustegtwo} }}{\card{ \ClustEst{\clusteg} } + \card{ \ClustEst{\clustegtwo} }} },
            \end{align}
            where we use estimates of the intracommunity edge density $\pinest$, and the intercommunity edge density $\pinterclustestnotime{\clusteg}{\clustegtwo}$.
            Condition \eqref{eqn:algdetectmerge} is similar to \eqref{eqn:mergethresh}, except with an additional scaling parameter $\DetectThresh$ in the right hand side.
            If $\DetectThresh\!=\!1$
            then a shrinking density gap $\pinest-\pinterclustestnotime{\clusteg}{\clustegtwo}$ causes erratic behavior during node inference, resulting in nodes incorrectly being moved between the pair of merging communities.
            This in turn corrupts the estimates $\pinest,\pinterclustestnotime{\clusteg}{\clustegtwo}$.
            We set $\DetectThresh\!=\!2$ to trigger the merge earlier and avoid this issue.
            
            \textit{Complexity:}
            Constructing the estimates takes $\bigO \paren{ \clustsketchsizedef^2 }$ time, whereas checking the merge condition for all pairs takes $\bigO \paren{ \NumClustSnapEst{\snaptime}^2}$ time.
        \end{alglevel}
        \label{step:mergedetect}

        \item
        \begin{alglevel}{\alglevelone}
            \textbf{Build estimate $\ClustSnapSetEst{\snaptime}$ using results of Steps~\ref{step:birthdetect}-\ref{step:mergedetect}}.
        \end{alglevel}
        \label{step:endloop}

\end{enumerate}

\noindent \textit{Output:} Partitions $\ClustSnapSetEst{\snaptime}, \snaptime=0,1,\dots$

\bigskip

Suppose that $\MaxClust$ and $\MaxNodes$ are the maximum number of communities and nodes, respectively, in any given snapshot.
We furthermore assume that $\initialsketchsize$ is at most $\MaxClust \clustsketchsizedef$.
Then, the computational complexity for estimating a single partition $\ClustSnapSetEst{\snaptime}$ at time $t \!\ge\! 0$
is $\bigO \paren{ \MaxClust \clustsketchsizedef (\MaxClust^2 \clustsketchsizedef^2 \!+\! \MaxNodes) }$.
For the first iteration, this is the time required for executing step~\ref{step:initclust}, whereas for each subsequent iteration, this is the total time required to execute steps~\ref{step:loopstart}-\ref{step:endloop}.
Contrast this with clustering the full snapshot graph, which is $\bigO \paren{ \MaxNodes^3 }$ for each iteration.
If $\MaxClust \!\ll\! \MaxNodes$ and we use a small sketch, this results in an order-wise improvement in complexity.

\section{Analysis} \label{sec:analysis}

In this section, we provide performance guarantees for the proposed algorithm, as well as guidelines for setting sketch size.
To simplify analysis we take the sketch to be balanced at all time steps, i.e., $\card{ \ClustSnapSketchEst{\clusteg}{\snaptime}} \!=\! \clustsketchsizedef$ for each  community $\clusteg$ and time $\snaptime$.
Furthermore, we suppose that the graph at the previous time step has been correctly clustered, i.e., $\ClustSnapSetEst{\snaptimeminusone}\!=\!\ClustSnapSet{\snaptimeminusone}$.
Unless otherwise specified, it is assumed that $\pinterclust{\clusteg}{\clustegtwo}{\snaptime} \!=\! \pout$ for any two communities $\clusteg$ and $\clustegtwo$.
The average degree of such a snapshot with $\NumClust$ communities is
$\AvgDeg = \clustsketchsizedef (\pin + \NumClust \, \pout)$.
The following approximation is made to provide clearer results.
\begin{assumption} \label{as:binomapprox}
Each of $\clustsimdyn{\nodeeg}{\clusteg}{\snaptime}$, $\pinest$, and $\pinterclustestnotime{\clusteg}{\clustegtwo}$ is well approximated by a normal random variable having the same mean and variance.
\end{assumption}
This assumption follows from the fact that the listed variables are driven by binomial random variables.
The underlying distributions of these random variables will generally have enough symmetry to be well approximated by normal distributions \cite{devore2012modern}.
More details are provided in \aref{sec:proofs}.

We now provide definitions used in this section.
Denote by $\norminv{\cdot}$ the inverse cumulative distribution function of the standard normal distribution. Specifically, given a standard normal random variable $Z$ and probability $\probofsuccess$, we have $\prob\paren{Z \le \norminv{\probofsuccess}} \!=\! \probofsuccess$.
Consider a graph with $\NumNodes$ nodes and
$\NumClust$ equal-sized communities 
$\ClustSet = \SetDef{\Clust{\clustiter}}{\clustiter \in \intsupto{\NumClust}}$.
The agreement with an estimated community partition $\ClustSetEst = \SetDef{\ClustEst{\clustiter}}{\clustiter \in \intsupto{\NumClust}}$
is defined as \cite{JMLR:v18:16-480}
\begin{align} \label{eqn:agree}
    \agree{\ClustSet}{\ClustSetEst} =
        \frac{1}{\NumNodes}
        \max_{\pi} \sum_{\clustiter = 1}^{\NumClust}
        \abs{ \Clust{\clustiter} \cap \ClustEst{\pi(\clustiter)} },
\end{align}
where $\pi$ ranges over the permutations on $\NumClust$ elements (this permutation is necessary since the community indices may be ordered arbitrarily).
Exact recovery is solved by an algorithm if it produces community estimates such that $\prob\paren{\agree{\ClustSet}{\ClustSetEst} = 1} \rightarrow 1$ as $\NumNodes \rightarrow \infty$.
In this section, we use 20 trials for each experiment.
Detailed derivations for the results in this section are deferred to \aref{sec:proofs}.

\subsection{Estimating communities in the initial sketch}
\label{sec:inititalsketch}

This section provides guidelines for choosing initial sketch size.
For simplicity, we consider the symmetric case in which every community has $\clustsizedef$ nodes.
Suppose that an initial sketch
has been constructed by sampling $\initialsketchsize$ nodes from $\GraphSnap{0}$.
If $a = \frac{\pin \initialsketchsize}{\ln(\initialsketchsize)}$ and $b = \frac{\pout \Nsamp}{\ln( \initialsketchsize)}$ are held constant, then exact recovery of the planted partition is efficiently solvable in the initial sketch provided $(\sqrt{a} \!-\! \sqrt{b})^2 \!>\! \NumClustSnap{0}$ \cite{JMLR:v18:16-480}.
Although this bound is only exact in the limit, it can still be used to estimate values of $\initialsketchsize$ for which agreement will remain high.
Specifically, for fixed $\initialsketchsize$, $\pin$, $\pout$, this bound becomes
\begin{align} \label{eqn:BoundExactRecovery}
        \frac{\initialsketchsize}{\ln(\initialsketchsize)}
        > \NumClustSnap{0} (\sqrt{\pin} - \sqrt{\pout})^{-2}.
\end{align}
Either a small density gap $\pin \!-\! \pout$ or a large number of communities $\NumClustSnap{0}$ can make the initial estimate unreliable.
These issues can be mitigated by increasing the sketch size.

\begin{figure}
\centering
\includegraphics[scale=1]{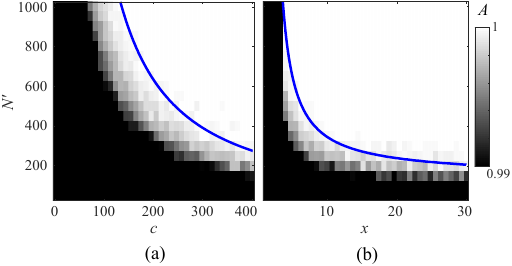}
\caption{
Plot of agreement for community estimates produced by $\NBAlg{\GraphSketch}$, where sketch graph $\GraphSketch$ is produced by randomly sampling $N'$ nodes from the full graph.
The edge densities are determined from average degree
$\AvgDeg$ and ratio $x = \pin / \pout$. Plots are shown for
(a) varying $c$ with $x=5$ and
(b) varying $x$ with $c=200$.
}
\label{fig:suffinitial}
\end{figure}

We demonstrate the efficacy of \eqref{eqn:BoundExactRecovery} in \fref{fig:suffinitial}.
We produce a sketch from a graph with two communities of size $\clustsizedef\!=\!2500$,
and plot agreement between the estimated and planted communities.
The blue line indicates the boundary of \eqref{eqn:BoundExactRecovery}.
Indeed, the agreement remains high (exceeding $0.998$) whenever this condition holds.

We note that if the initial snapshot is imbalanced, i.e., with communities of different size, the \acs{SPIN} (\textit{\acl{SPIN}}) sampling method \cite{Beckus_MLSP:19} may be used in place of uniform random sampling.
This method can improve the success rate by sampling more uniformly across communities.

\subsection{Birth detection}
\label{sec:birthanalysis}

Suppose that one or more new communities are born at time $\snaptime$, and take a node $i \!\in\! \NodeBirth$ which belongs to one of these communities.
Let $\pineststddev \!=\! \sqrt{\frac{2 \pin(1-\pin)}{\NumClustSnap{t} \clustsketchsizedef(\clustsketchsizedef\!-\!1)}}$
be the standard deviation of estimator $\pinest$.
Then, the probability that node $i$ is correctly identified as belonging to $\NodeBirth$ is at least $\probofsuccess$ if
\begin{align} \label{eqn:suffbirth}
    \clustsketchsizedef
    \ge
    \paren{
    \frac{\norminv{1\!-\!\birthprob}
          \sqrt{\pout(1-\pout)}
          +
          3 \sqrt{\pinestupperbound(1-\pinestlowerbound)}
         }
         {\pinestlowerbound - \pout}
    }^2
\end{align}
where
$\birthprob=\paren{1-\probofsuccess}/\paren{2(\NumClustSnap{t}-1)}$
and
\begin{subequations} \label{eqn:pinestupperlower}
\begin{align}
    \pinestupperlowerbound
    &= \pin \pm \norminv{1-\frac{\birthprob}{2}} \pineststddev.
\end{align}
\end{subequations}
Note that the sufficient number of samples in \eqref{eqn:suffbirth} is independent of community size in the full graph.
This allows for the detection of new communities even when they are of a very small size.
This advantage is illustrated further in the numerical results of \sref{sec:scalability}.

\begin{figure}
\centering
\includegraphics[scale=1]{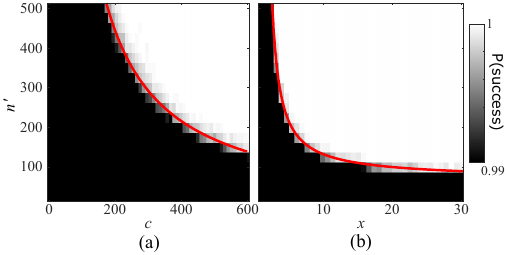}
\caption{
Empirical estimate of the probability of successfully detecting a node joining a new community.
The edge densities are determined from average degree
$\AvgDeg$ and ratio $x = \pin / \pout$. Plots are shown for
(a) varying average degree $\AvgDeg$ with $x=5$ and
(b) varying $x$ with $c=200$.
}
\label{fig:suffbirth}
\end{figure}

\fref{fig:suffbirth} shows results in which a new community with 500 nodes joins a graph containing two existing communities of size $\clustsizedef\!=\!2500$.
The plots indicate the fraction of nodes in $\NodeBirth$ which are correctly identified as belonging to the newborn community.
The red line shows the boundary of condition \eqref{eqn:suffbirth} with $\probofsuccess\!=\!0.99$, and shows excellent agreement with the numerical results.

As the density gap $\pin \!-\! \pout$ shrinks, a larger sketch will be required to reliably detect which nodes belong to newborn communities.
In fact, as $\clustsketchsizedef \!\rightarrow\! \infty$, we have $\pineststddev \!\rightarrow\! 0$ such that $\pinestupperlowerbound \rightarrow \pinest$, and the right side of \eqref{eqn:suffbirth} converges to
\begin{align} \label{eqn:suffbirthconverge}
    \paren{
    \frac{\norminv{1\!-\!\birthprob}
          \sqrt{\pout(1\!-\!\pout)}
          +
          3 \sqrt{\pin(1\!-\!\pin)}
         }
         {\pin \!-\! \pout}
    }^2.
\end{align}
In this regime, the denominator depends solely on the square of the density gap.

\subsection{Inferring community membership}
\label{sec:inferanalysis}

We next consider the required sketch size to successfully infer community membership of individual nodes using \eqref{eqn:estclust}.
Define $\nummovednodessketch{\clusteg}{\clustegtwo} =
    \card{ \movednodesetsnap{\clusteg}{\clustegtwo}{\snaptime} \cap \SketchIdxSnap{\snaptimeminusone} }$,
i.e., the number of nodes in the sketch that are moving from $\clusteg$ to $\clustegtwo$.
The analysis here will use the following simplification.
\begin{assumption} \label{as:movesketchexpect}
In place of random variable $\nummovednodessketch{\clusteg}{\clustegtwo}$, we use its expected value
$\expect{\nummovednodessketch{\clusteg}{\clustegtwo}} = 
\clustsketchsizedef \card{\movednodeset{\clusteg}{\clustegtwo}} / \card{\ClustSnap{\clusteg}{\snaptimeminusone}}$.
\end{assumption}
\noindent We denote the minimum community size in a given snapshot by $\clustsizemin = \min_{1 \le \clustiter \le \NumClustSnap{\snaptime}}
\card{ \ClustSnap{\clustiter}{\snaptime} }$.

Suppose that at most $\maxmovednodes$ nodes move between any two pairs of communities, i.e.,  $\card{\movednodesetsnap{\clustiter}{\clustitertwo}{\snaptime}} \le \maxmovednodes$ for $1 \le \clustiter, \clustitertwo \le \NumClustSnap{\snaptime}$. 
Then, \eqref{eqn:estclust} correctly  identifies the community of a given node $\nodeeg \notin \SketchIdxSnap{\snaptime}$ with probability $\ge \probofsuccess$ provided that
\begin{align} \label{eqn:suffmoveinfer}
    \SketchClustSize
    \ge
    \infersuffnormalvarbound
    \bracket{
    \frac{\norminv{1-\frac{1-\probofsuccess}{\NumClustSnap{\snaptime}-1}}}
         {\infersuffnormalmeanbound}
    }^2
\end{align}
where
\begin{align}
    \infersuffnormalmeanbound
    =& (\pin-\pout) \paren{1- \frac{\maxmovednodes \, \NumClustSnap{\snaptime}}{\clustsizemin}},
\\
    \infersuffnormalvarbound
    =& \pin (1-\pin) + \pout (1-\pout)
\\
    &+ \maxmovednodes \frac{\pin (1-\pin) - \pout (1-\pout)}
    {\clustsizemin}.
\end{align}
Variable $\infersuffnormalmeanbound$ serves as a lower bound on the expected value of
$\clustsimdyn{\nodeeg}{\clusteg}{\snaptime}-\clustsimdyn{\nodeeg}{\clustiter}{\snaptime}$ for $\clustiter \ne \clusteg$,
whereas $\frac{\infersuffnormalvarbound}{\clustsketchsizedef}$ serves as an upper bound on the standard deviation.
Both a small density gap $\pin \!-\! \pout$ and a large number of moving nodes $\maxmovednodes$ can make inference less reliable. In these situations, an increased sketch size will be required to keep the probability of misclassification low.

\begin{figure}
\centering
\includegraphics[scale=1]{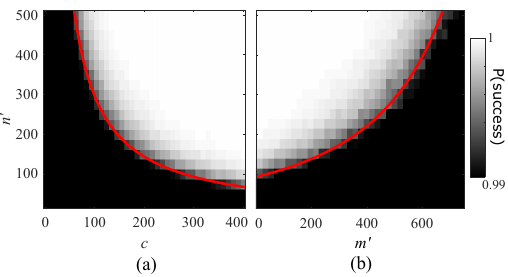}
\caption{
Plots show empirical estimate of success probability of \eqref{eqn:estclust} over $10^5$ trials.
Sketch graph $\GraphSketch$ is produced by randomly sampling $N'$ nodes from a network with the same parameters as in \fref{fig:suffinitial}.
Let $\nummovednodessketch{1}{2}=\nummovednodessketch{2}{1}=m'$, i.e., $m'$ nodes move from one community to the other at each time step, and $\pin\!=\!5\pout$.
(a) Varying average degree $c$ with $m'=0$.
(b) Varying $m'$ with $c=300$.}
\label{fig:suffinfer}
\end{figure}

\fref{fig:suffinfer} shows the inference success rate of \eqref{eqn:estclust} for a network containing two communities of size $\clustsizedef\!=\!2500$.
In \fref{fig:suffinfer}(a), $\maxmovednodes=0$ (no nodes move between communities), whereas \fref{fig:suffinfer}(b) varies $\maxmovednodes$.
The red lines indicate the boundary of \eqref{eqn:suffmoveinfer} with $\probofsuccess=0.99$, and show excellent agreement with the numerical results.
As $\maxmovednodes \rightarrow 0$,
this boundary converges to
\begin{align} \label{eqn:suffrhsinfer}
    \bracket{\norminv{1-\frac{1-\probofsuccess}{\NumClustSnap{\snaptime}-1}}}^2
    \paren{ \frac{\pin(1-\pin)+\pout(1-\pout)}{(\pin-\pout)^2} },
\end{align}
which is independent of community size in the full graph.
This advantage will be illustrated further in the numerical results of \sref{sec:scalability}.
In this case, the primary driver of performance becomes the density gap $\pin-\pout$.

\subsection{Split detection}
\label{sec:splitanalysis}

Consider a network with a single community undergoing a split into two equal-sized communities $\clusteg$ and $\clustegtwo$.
An important consideration is the smallest value of $\pinterclustnotime{\clusteg}{\clustegtwo}$ at which the communities will be considered split according to \eqref{eqn:splitdecision}.
Using a similar argument as for the initial sketch, in practice we may use the exact recovery limit to approximate this lower bound.
Likewise, we can use the asymptotic detectability threshold as an approximate upper bound.
Following this line of reasoning, it is likely that the split will be detected for some $\pinterclustnotime{\clusteg}{\clustegtwo}$ bounded according to
\begin{align} \label{eqn:SplitBoundExactRecovery}
&\pin
+ \frac{1}{2 \clustsketchsizedef}
- \sqrt{\frac{2 \pin}{\clustsketchsizedef} + \frac{1}{4 \clustsketchsizedef^2}}
\nn\\
&>\pinterclustnotime{\clusteg}{\clustegtwo}
>
\paren{\sqrt{\pin} - \sqrt{\frac{\ln(2 \clustsketchsizedef)}{\clustsketchsizedef}}}^2.
\end{align}
Increased sketch size will tend to allow earlier detection of the split, i.e., for smaller values of $\pin-\pinterclustnotime{\clusteg}{\clustegtwo}$.

\begin{figure}
\centering
\includegraphics[scale=1]{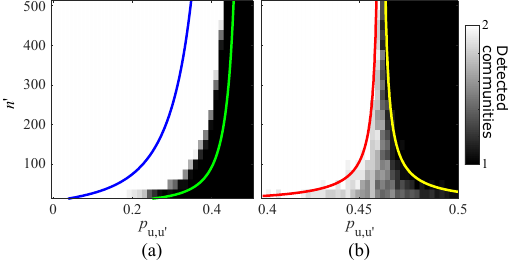}
\caption{
Detected number of communities for a pair of communities undergoing the merge-split process.
Detection is based on
(a) split condition \eqref{eqn:splitdecision}
and
(b) merge condition \eqref{eqn:algdetectmerge}.
}
\label{fig:suffmergesplit}
\end{figure}

\fref{fig:suffmergesplit}(a) shows a plot of the estimated number of communities from  \eqref{eqn:splitdecision} for a sketch with two communities containing $\clustsketchsizedef$ nodes each
(the detected number of communities is 2 if the condition holds, and 1 otherwise).
Along the horizontal axis, we vary $\pinterclustnotime{\clusteg}{\clustegtwo}$ within $[0,\pin]$, where $\pin\!=\!0.5$.
The blue and green lines show the lower and upper bounds in \eqref{eqn:SplitBoundExactRecovery}, respectively.
The split is indeed detected for a value of $\pinterclustnotime{\clusteg}{\clustegtwo}$ within these bounds.

\subsection{Merge detection}
\label{sec:mergeanalysis}

Finally, suppose that two equal-sized communities $\clusteg, \clustegtwo$ are merging into one community.
We consider the value of $\pinterclustnotime{\clusteg}{\clustegtwo}$ at which condition \eqref{eqn:algdetectmerge} detects a merge.
This condition will hold with probability $\ge \probofsuccess$ if
\begin{align} \label{eqn:suffmergeq}
    \pinterclustnotime{\clusteg}{\clustegtwo}
    \ge
    \pin
    \!+\! \norminv{1\!-\!\frac{1\!-\!\probofsuccess}{2}} \mergediffsumstd
    \!+\! \frac{\DetectThresh^2}{2\clustsizedef} - \DetectThresh \sqrt{\frac{2\pin}{\clustsizedef} + \frac{\DetectThresh^2}{\clustsizedef^2}},
\end{align}
where we use the the standard deviation of $\pinest \pm \pinterclustestnotime{\clusteg}{\clustegtwo}$,
\begin{align}
    \mergediffsumstd
    = \sqrt{
    \frac{2 \pin(1-\pin)}{\NumClustSnap{\snaptime} \clustsketchsizedef(\clustsketchsizedef-1)}
    + \frac{\pinterclustnotime{\clusteg}{\clustegtwo}(1-\pinterclustnotime{\clusteg}{\clustegtwo})}{(\clustsketchsizedef)^2}
    }.
\end{align}
However, it is also important to consider when \eqref{eqn:algdetectmerge} reliably identifies the communities as being split.
This occurs with probability $\ge \probofsuccess$ if
\begin{align} \label{eqn:suffsplitdetq}
    \pinterclustnotime{\clusteg}{\clustegtwo}
    \le
    \pin
    \!-\! \norminv{1\!-\!\frac{1\!-\!\probofsuccess}{2}} \mergediffsumstd
    \!+\! \frac{\DetectThresh^2}{2\clustsizedef} - \DetectThresh \sqrt{\frac{2\pin}{\clustsizedef} + \frac{\DetectThresh^2}{\clustsizedef^2}}.
\end{align}

To illustrate the significance of bounds \eqref{eqn:suffmergeq} and \eqref{eqn:suffsplitdetq}, \fref{fig:suffmergesplit}(b) shows the detected number of communities for a pair of communities with $\clustsizedef\!=\!2500$ nodes each (the detected number of communities is 1 if \eqref{eqn:algdetectmerge} holds, and 2 otherwise).
The red line indicates the boundary of \eqref{eqn:suffmergeq}, and the yellow line indicates the boundary of \eqref{eqn:suffsplitdetq}, for $\probofsuccess=0.9$.
When $\pinterclustnotime{\clusteg}{\clustegtwo}$ falls in the gap between these bounds, the detection tends to be unreliable.
However, the size of this gap can be reduced by using larger sketch sizes to drive down the standard deviation $\mergediffsumstd$.

\section{Numerical Results} \label{sec:results}

We compare against four algorithms from the literature, each of which uses a different means for carrying forward information from one snapshot to the next.
First, we use the classic Bayesian approach found in Yang \textit{et al.} \cite{Yang2011}.
Second, we run the algorithm of Dinh \textit{et al.} \cite{5403845}.
This algorithm uses a sketch-like concept by consolidating known communities into ``supernodes'' within a weighted graph.
These supernodes are then incorporated into the next snapshot.
Third, we use \acs{ESPRA} (\textit{\acl{ESPRA}}), which is based on structural perturbation theory \cite{Wang_2017}.
This algorithm defines an objective function which explicitly balances two similarities: one which encourages temporal smoothness across snapshots, and one that takes into account only the community structure in the latest snapshot.
Lastly, we independently cluster each snapshot as described in such works as \cite{Hopcroft5249,Aynaud2013}.
This algorithm, referred to here as \StdInd/, estimates the communities in the current snapshot using $\NBAlgnoparams$, and then matches the estimates in adjacent snapshots using the Jaccard similarity coefficient \cite{doi:10.1111/j.1469-8137.1912.tb05611.x}.
Although $\NBAlgnoparams$ performs optimally in certain regimes, the main weakness of \StdInd/ is that it completely ignores information from the previous snapshot when clustering the current snapshot.

Further details regarding these algorithms are provided in \aref{sec:comparisonalgorithms}.
Unless otherwise specified, all plots show an average over 20 independent runs.
We set the initial sketch size to $\initialsketchsize=\NumClustSnap{0} \clustsketchsizedef$.

\subsection{Performance with small clusters}

We first consider the performance of the proposed algorithm in the presence of small communities.
We use normalized agreement to compare the planted communities
$\ClustSet=\SetDef{\Clust{\clustiter}}{\clustiter \in \intsupto{\NumClust}}$
and estimated communities
$\ClustSetEst=\SetDef{\ClustEst{\clustiter}}{\clustiter \in \intsupto{\NumClustEst}}$.
Sets $\ClustSet$ and $\ClustSetEst$ are padded with empty communities such that $| \ClustSet |\!=\!| \ClustSetEst |$.
Then, normalized agreement is defined as \cite{JMLR:v18:16-480}
\begin{align} \label{eqn:normagree}
    \normagree =
        \frac{1}{\NumClust}
        \max_{\pi}
        \sum_{\stackrel{\clustiter=1}
                       {\card{ \Clust{\clustiter} } > 0}}
            ^{\NumClust}
        \frac{ \card{ \Clust{\clustiter} \cap \ClustEst{\pi(\clustiter)} } }
        { \card{ \Clust{\clustiter} } },
\end{align}
where $\pi$ ranges over the permutations on $\NumClust$ elements.
The normalized agreement for the snapshot at time $t$ is denoted $\normagreesnap{t}$.
Unlike the agreement metric defined earlier in \eqref{eqn:agree}, normalized agreement proves useful for quantifying performance in the presence of small clusters, since each community constitutes a fraction $\frac{1}{\NumClustSnap{\snaptime}}$ of the normalized agreement, regardless of community size.

For summarizing the overall deviation in the actual and estimate communities for a snapshot sequence, we use the average-squared error
\begin{align}
    \squarederrnormagree  = \frac{1}{\period} \sum_{\snaptime=1}^\period \bracket{ 1-\normagreesnap{\snaptime} }^2,
\end{align}
where $\period$ is the total number of snapshots.

\subsubsection{Grow-shrink benchmark}
\label{sec:smallgrowshrink}

\begin{figure}
\centering
\includegraphics[scale=1]{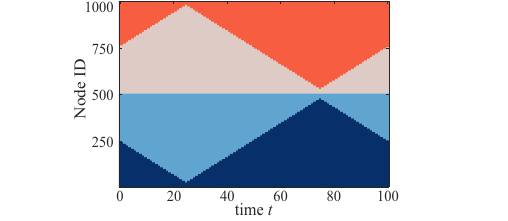}
\caption{
Planted partitions for a double-stacked version of the grow-shrink benchmark, with $\clustsizedef\!=\!250,\growshrinkf\!=\!0.95$, $\NumClust\!=\!4,\pin\!=\!0.4,\pout\!=\!0.1$.
}
\label{fig:SmallClusterPhaseGrowShrink}
\end{figure}

\begin{figure*}
\centering
\includegraphics[scale=1]{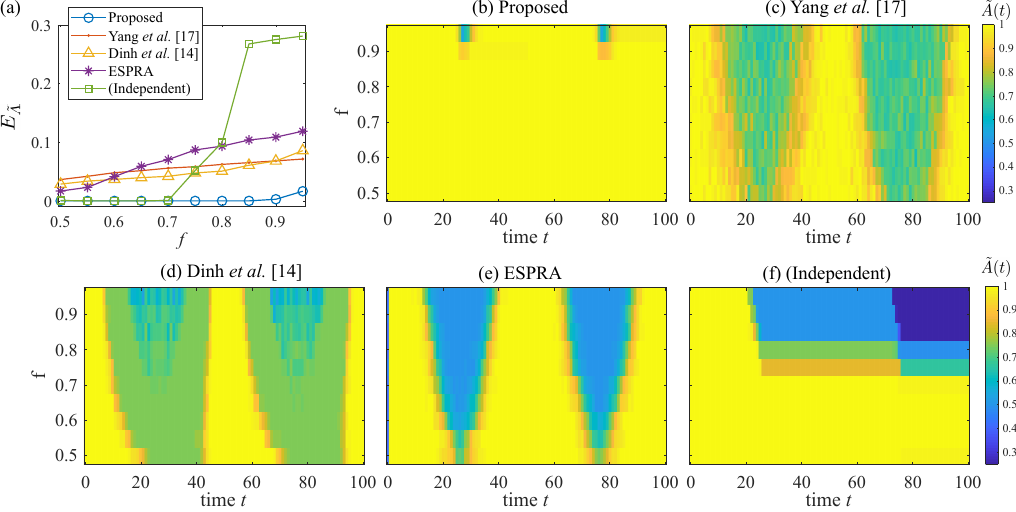}
\caption{
Results for varying $\growshrinkf$ in the grow-shrink example in \fref{fig:SmallClusterPhaseGrowShrink}.
For the proposed algorithm we set $\clustsketchsizedef\!=\!50$.
Plot of $\squarederrnormagree$ is shown in (a).
Panels (b) through (f) show ensemble averages of $\normagreesnap{\snaptime}$ as a function of time along the horizontal axis, and $\growshrinkf$ along the vertical axis for each algorithm.
}
\label{fig:GrowShrinkVaryf}
\end{figure*}

We use two concurrent instances of the grow-shrink benchmark with phase $\benchphase\!=\!0$ for the first instance, and $\benchphase\!=\!\benchperiod/2$ for the second instance.
Figure~\ref{fig:SmallClusterPhaseGrowShrink} shows planted partitions for an example with $\growshrinkf\!=\!0.95$.
The community detection results are shown for all algorithms in \fref{fig:GrowShrinkVaryf}(a),
where the value of $\squarederrnormagree$ is plotted as a function of $\growshrinkf$.
The proposed algorithm has $\squarederrnormagree\!<\!0.02$ for all values of $\growshrinkf$,
whereas the other algorithms exhibit significantly larger values of $\squarederrnormagree$ especially for larger $\growshrinkf$.

To gain further insight into the behavior of the algorithms, we plot a heat map of $\normagreesnap{\snaptime}$ for each algorithm in \fref{fig:GrowShrinkVaryf}(b)-(f), with $\growshrinkf$ varied along the vertical axis and time $\snaptime$ along the horizontal.
Increasing values of $\growshrinkf$ result in smaller communities at times $\snaptime\!=\!\benchperiod/4$ and $\snaptime\!=\!3 \benchperiod/4$ when the graph is most imbalanced.
It is exactly around these times that the algorithms tend to perform worst.
\StdInd/ often loses track of the small clusters at $\snaptime\!=\!\benchperiod/4$ and $\snaptime\!=\!3\benchperiod/4$, resulting in a merge of communities and a sharp drop in agreement.
The algorithm is not capable of detecting splits, and so does not recover.

\subsubsection{Birth-death benchmark}
\label{sec:smallbirthdeath}

\begin{figure}
\centering
\includegraphics[scale=1]{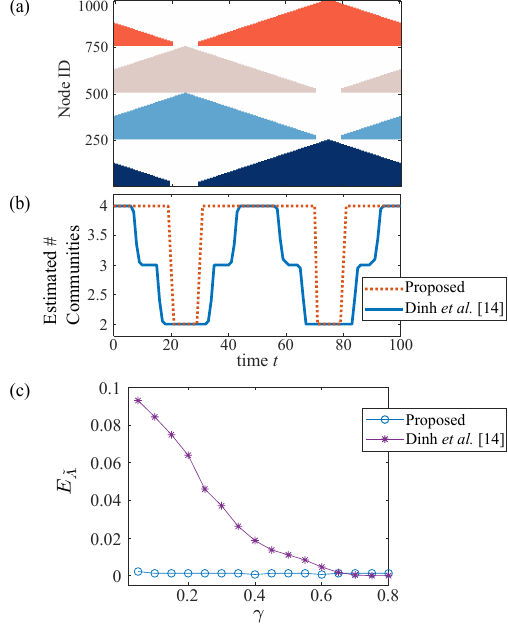}
\caption{
(a) Planted partitions for a double-stacked version of the birth-death benchmark, with $\birthdeathgap\!=\!0.1$.
For the first instance, we use phase shift $\benchphase\!=\!0$, whereas for the second instance we use $\benchphase\!=\!\benchperiod/2$.
Both instances have parameters $\NumClust\!=\!4,\pin\!=\!0.5,\pout\!=\!0.05$.
We set $\clustsketchsizedef\!=\!50$, which leads to a maximum sketch size of 200 nodes.
(b) Ensemble average of number of communities estimated by algorithm plot as a function of time.
(c) Squared error of normalized agreement $\squarederrnormagree$ is shown for varying $\birthdeathgap$. 
}
\label{fig:SmallClusterPhaseBirthDeath}
\end{figure}

We now present analogous examples for the birth-death benchmark.
One means for producing small clusters is by using small values of $\birthdeathgap$, such that each community is small immediately after birth and before death.
An example is shown in \fref{fig:SmallClusterPhaseBirthDeath}(a), with $\birthdeathgap\!=\!0.1$.
We execute the algorithms and show the estimated number of communities for each snapshot in \fref{fig:SmallClusterPhaseBirthDeath}(b).
The algorithm of \Dinh/ tends to absorb small communities into the larger communities, as exhibited by the drop in estimated number of communities after birth and before death.
Meanwhile, the proposed algorithm provides a near-perfect estimate.
We expand on this example by plotting $\squarederrnormagree$ as a function of $\birthdeathgap$ in \fref{fig:SmallClusterPhaseBirthDeath}(c).
The proposed algorithm has $\squarederrnormagree\!<\!0.003$ for all values of $\birthdeathgap$.
We omit \StdInd/, Yang \textit{et al.} \Yang/, and \acs{ESPRA} as they cannot handle graphs of changing size, nor new communities.

\subsection{Scalability}
\label{sec:scalability}

To demonstrate the scalability of the proposed algorithm, let us consider the minimum community size over all snapshots
\begin{align}
\nminscaleresult=
\min_{1 \le \snaptime \le \period}
\,\,
\min_{1 \le \clustiter \le \NumClustSnap{\snaptime}}
\card{ \ClustSnap{\clustiter}{\snaptime} }.
\end{align}
We run the grow-shrink benchmark using the same parameters as in \sref{sec:smallgrowshrink}, except with $\growshrinkf \!=\! 1-\frac{\clustsizedef}{\nminscaleresult}$ such that the minimum cluster sizes are fixed at $\nminscaleresult=200$.
\tabref{tab:scalability} shows the value of $\squarederrnormagree$ as a function of $\clustsizedef$, averaged over five trials.
There is a small increase in $\squarederrnormagree$ as $\clustsizedef$ increases, due to a corresponding increase in the number of moving nodes (as described in \sref{sec:inferanalysis}).
Nonetheless, $\squarederrnormagree$ remains below $5.5\!\times\!10^{-5}$ despite a dramatic increase in imbalance of the full graph, and despite the fact that the sketch size remains fixed.
\begin{table}[t!]
  \centering
  \caption{\bf Scalability of proposed algorithm.}
  \begin{tabular}{c c c c}
  \hline
  Benchmark & $\clustsizedef$ & $\squarederrnormagree$ & 
        \begin{tabular}{c}
           Runtime \\
           (normalized)
        \end{tabular} \\
  \hline
    \multirow{4}{*}{
        \begin{tabular}{c}
           Grow-Shrink \\
           ($\nminscaleresult=200$)
        \end{tabular}
    }
          & $250$  & $3.0\!\times\!10^{-7}$ & $1$ \\
          & $1000$ & $7.2\!\times\!10^{-7}$ & $1.2$ \\
          & $2000$ & $4.6\!\times\!10^{-6}$ & $2.2$ \\
          & $3000$ & $5.5\!\times\!10^{-5}$ & $3.7$ \\
    \hline
    \multirow{4}{*}{
        \begin{tabular}{c}
           Birth-Death \\
           ($\nminscaleresult=20$)
        \end{tabular}
    }
          & $250$  & $1.6\!\times\!10^{-5}$ & $1$ \\
          & $1000$ & $2.4\!\times\!10^{-5}$ & $1.2$ \\
          & $2000$ & $2.6\!\times\!10^{-7}$ & $1.6$ \\
          & $3000$ & $2.6\!\times\!10^{-7}$ & $1.7$ \\
    \hline
  \end{tabular}
  \label{tab:scalability}
\end{table}

Likewise, we run the birth-death benchmark with the parameters of \sref{sec:smallbirthdeath}, but with $\birthdeathgap = 2\,\nminscaleresult/\clustsizedef$ such that $\nminscaleresult\!=\!20$ regardless of graph size.
The smallest community size is attained immediately before death and after birth.
The results are shown in \tabref{tab:scalability}.
Unlike the results for the grow-shrink benchmark, there is no increase in $\squarederrnormagree$.
This is consistent with the analysis in \sref{sec:birthanalysis}, which showed no dependence on community size in the full graph.

\begin{figure}
\centering
\includegraphics[scale=1]{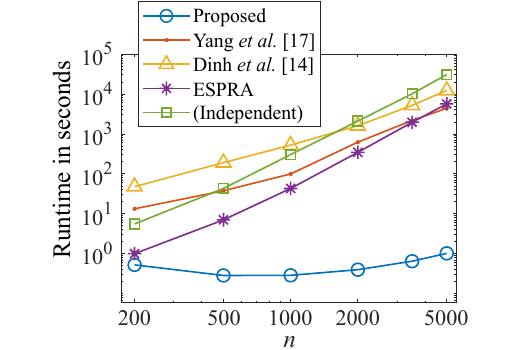}
\caption{
Timing results for the grow-shrink benchmark with $\NumClust\!=\!2,\pin\!=\!0.5,\pout\!=\!0.05,\growshrinkf\!=\!0.5$.
For the proposed algorithm we set $\clustsketchsizedef\!=\!100$, leading to a total sketch size of 200 nodes.
Due to the large runtimes in the four algorithms we compare against, only ten iterations of the algorithms are performed, and time is averaged over five trials.
Note that logarithmic scales are used for both axes.
All algorithms had perfect community estimates for all network sizes, except for \acs{ESPRA} which still had less than 1\% misclassified nodes per snapshot.
}
\label{fig:timings}
\end{figure}

For both benchmarks, \tabref{tab:scalability} shows only a sub-linear increase in runtime as community size $\clustsizedef$ increases, owing to the fixed sketch size.
To expand on this result, we run all of the algorithms on the grow-shrink benchmark, and show the results as a function of $\clustsizedef$ in \fref{fig:timings}.
As expected, the proposed algorithm finishes very fast, in under two seconds for all cases.
On the other hand, algorithms \Yang/, \acs{ESPRA}, and \StdInd/ all cluster the full graph, and therefore scale super-linearly with network size.
Although \Dinh/ clusters a graph of reduced size at each time step, nodes having changed edges are left as singleton nodes.
In this example, the large number of edge changes forces a correspondingly large number of nodes to remain singletons, thus requiring the static clustering step to operate on large networks.

\subsection{Merge-split detection}
\label{sec:mergesplitdetect}

\begin{figure}
\centering
\includegraphics[scale=1]{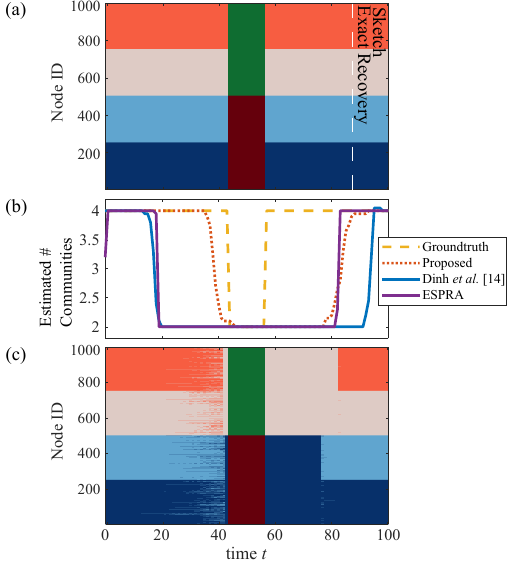}
\caption{
(a) Planted partitions for a double-stacked version of the merge-split benchmark.
The parameters of the model are $\NumClust\!=\!4, n\!=\!250,\pin\!=\!0.5,\pout\!=\!0.05$.
We set $\clustsketchsizedef\!=\!50$ for the proposed algorithm.
This results in a sketch size of 100 in the merged state, and 200 in the split state.
The exact recovery limit for the sketch, based on \eqref{eqn:BoundExactRecovery}, is shown as a dashed white vertical line.
(b) The estimated number of communities at each time step for the proposed algorithms, as well for \Dinh/ and \acs{ESPRA}.
(c) The estimated partitions for the proposed algorithm.
}
\label{fig:MergeSplit}
\end{figure}

We next execute the algorithms on the merge-split benchmark, with two concurrent instances as shown in \fref{fig:MergeSplit}(a).
The plot in \fref{fig:MergeSplit}(b) shows the estimated number of communities as a function of time for the proposed algorithm, as well as for \Dinh/ and \acs{ESPRA}.
We omit \StdInd/ and \Yang/ as they cannot handle the merge and split processes.
Although the benchmark groundtruth undergoes an instantaneous transition between merged and split states, the network itself gradually interpolates between these states.
This discrepancy in timescales, along with the fact that the benchmark sets the transition at the theoretical detectability limit, means we cannot expect the estimated partitions to exactly match the planted partitions.
Indeed, all three algorithms overestimate the span of time during which the communities are merged.

The estimates of the proposed algorithm are shown in \fref{fig:MergeSplit}(c),
where we can see that nodes start being misclassified at $\snaptime\!=\!13$.
This is expected due to the shrinking gap between $\pin$ and the intercommunity edge densities, as described in \sref{sec:SummaryAlg}.
Nonetheless, the proposed algorithm detects the merge much closer to the benchmark's merge time than the other two algorithms.
We note that using larger values of $\DetectThresh$ in condition \eqref{eqn:algdetectmerge} will result in an earlier detection of the merge.
In this way, $\DetectThresh$ can act as a tuning parameter to adjust the sensitivity of the merge detection.

For studying the performance of the algorithms' split detection, we show the exact recovery limit for the sketch as a vertical white dashed line in \fref{fig:MergeSplit}(a).
The proposed algorithm detects the split close to this limit, although we point out that the detection could be shifted earlier by increasing the sketch size.
Despite clustering the full network, for which estimation should be easier, \acs{ESPRA} does not exceed the performance of the proposed algorithm, and \Dinh/ fares even worst.

\subsection{Mixed benchmark} \label{sec:resultssimultaneous}

\begin{figure}
\centering
\includegraphics[scale=1]{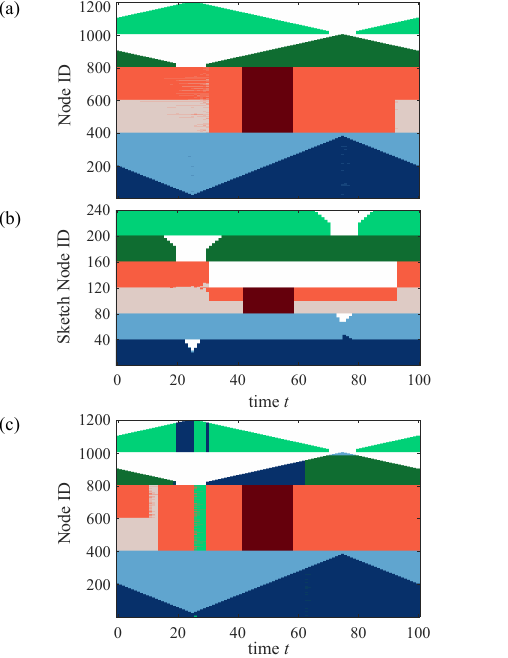}
\caption{
Results for mixed benchmark with $\pin\!=\!0.5,\pout\!=\!0.05$.
Estimated partitions produced by the proposed algorithm are shown for (a) the full network and (b) the sketches produced by the proposed algorithm.
The estimated partitions produced by \Dinh/ are shown in (c).
}
\label{fig:Mixed3Results}
\end{figure}

So far, our results have considered individual benchmarks in isolation.
We now run the proposed algorithm on the mixed benchmark from \fref{fig:birthdeathdiagram}(a), which has concurrent birth-death, grow-shrink and merge-split processes.
The partition estimates are shown in \fref{fig:Mixed3Results}(a).
Most of the mismatch occurs in the merge-split communities, which is consistent with our earlier results.

The  sketches produced by the proposed algorithm are shown in \fref{fig:Mixed3Results}(b).
The sketch nodes are sorted vertically according to their planted communities, with the color indicating the estimated community of the corresponding node.
The deviation from the ideal sketch in \fref{fig:birthdeathdiagram}(b) lies mostly inside the merge-split communities, due to the errors present in the estimates of the full graph.

The estimated partitions for \Dinh/ are presented in \fref{fig:Mixed3Results}(c).
As with the earlier results, \Dinh/ encounters difficulties in correctly identifying the small clusters in the grow-shrink and birth-death communities.

\section{Conclusion} \label{sec:conclusion}

This paper concerned a sketch-based approach for community detection in time-evolving networks.
We presented an \ac{SBM}-based model along with possible evolutionary processes which may occur within this model.
We then proposed sketch-based techniques for tracking these processes, as well as an algorithm incorporating these techniques to produce community estimates for concurrent processes.
We provided an analysis to guide the choice of sketch size, and generated numerical results comparing the proposed algorithm to full-scale community detection algorithms.

We conclude by briefly noting possible extensions.
First, an arbitrary community detection algorithm may be used in place of $\NBAlgnoparams$, provided that it can estimate the number of communities.
Second, a straightforward extension to the network model and algorithm would allow the intracommunity edge density $\pin$ to vary for each community.
Third, our approach is extendable to other graph models as well, for example the Degree Corrected \ac{SBM} (DCSBM) \cite{PhysRevE.83.016107}.
This can be accomplished by substituting a suitable sampling technique for constructing DCSBM sketches (e.g., the sampling method of \cite{9596377}), a similarity definition between nodes in the full network and the sketch communities, and an appropriate technique for determining when clusters split or merge.
\bigskip 

\begin{acknowledgments}
This work was supported by NSF CAREER Award CCF-1552497 and NSF Award CCF-2106339.
The University of Central Florida Advanced Research Computing Center provided computational resources that contributed to results reported herein.
\end{acknowledgments}

\appendix

\section{Static clustering}
\label{sec:staticclust}

We use the following algorithm to perform static clustering of graph $\Graph$ with $\NumNodes$ nodes.

\bigskip

$\NBAlg{\Graph}$
\begin{enumerate}[topsep=0pt,noitemsep,label={(\arabic*)},ref=\theenumi]

    \item Construct $\NBmatrixSurrogate$ from $\Graph$ using \eqref{eqn:NBsurrogate}.
    \label{step:staticstartclust}

    \item Calculate eigenvalues $\NBEig{1}, \NBEig{2}, \dots, \NBEig{\NumNodes}$ and corresponding eigenvectors of $\NBmatrixSurrogate$.
    
    \item Calculate $\NumClustEst$ as the maximum value of $i$ such that $\lambda_i > \sqrt{\lambda_1}$.
    \label{step:staticendclust}

    \item \AlgCmd{for} $i = 1 \dots \NumClustEst$ \AlgCmd{do}
    \label{step:staticstartsearch}
        \item
        \begin{alglevel}{\alglevelone}
            Build matrix $\bM$ from the $i$ normalized eigenvectors of $\NBmatrixSurrogate$ corresponding to eigenvalues $\EigVal{1}, \dots, \EigVal{i}$.
            Apply k-means clustering to $\bM$ to obtain community estimates $\ClustSetEst_i=\SetDef{\ClustEst{\clustiter}}{\clustiter \in \intsupto{i}}$
            We repeat 100 iterations with three random initializations and take the best result.
        \end{alglevel}
    
        \item
        \begin{alglevel}{\alglevelone}
            Calculate modularity $Q_i$ of $\Graph$ with partition $\ClustSetEst_i$.
            Modularity is defined as in \cite[Section IV]{PhysRevE.69.026113}.
        \end{alglevel}

    \item \AlgCmd{end for}

    \item $j \gets \argmax_{1 \le i \le \NumClustEst} \Modularity_i$
    \label{step:staticendsearch}
    
    \item Return estimate $\ClustSetEst_j$.
    
\end{enumerate}
\bigskip

Steps~\ref{step:staticstartclust}-\ref{step:staticendclust} estimate the number of communities $\NumClustEst$, and are as described in \cite{Krzakala20935}.
We find that adding Steps~\ref{step:staticstartsearch}-\ref{step:staticendsearch} provides a more reliable estimate of the number of communities.
These steps repeat k-means clustering, varying the number of clusters up to $\NumClustEst$, and then return the partition giving the highest modularity.

\begin{figure}
\centering
\includegraphics[scale=1]{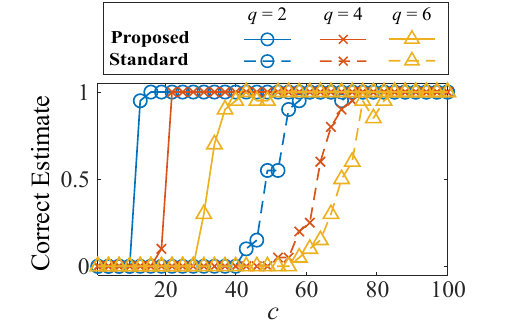}
\caption{
Results showing improved performance of the proposed static clustering algorithm afforded by the modularity-based heuristic in Steps~\ref{step:staticstartsearch}-\ref{step:staticendsearch}.
We generate a graph $\Graph$ with $800$ nodes divided into $\NumClustEst$ equal-sized communities, and $\pin=5\pout$.
}
\label{fig:StaticModularityPerformance}
\end{figure}

\fref{fig:StaticModularityPerformance} compares the proposed function $\NBAlgnoparams$ (solid lines), against the standard approach (dashed lines).
The standard approach only uses k-means to identify $\NumClustEst$ communities, rather than executing Steps~\ref{step:staticstartsearch}-\ref{step:staticendsearch}.
The plot shows the fraction of runs in which the estimated number of communities is correct, out of 20 runs. 
As can be seen, the proposed algorithm identifies the correct number of communities for much smaller values of average degree $\AvgDeg$.

The additional steps do not increase the complexity of the algorithm.
In particular, the time for Steps~\ref{step:staticstartsearch}-\ref{step:staticendsearch} is $\bigO(\NumNodes \NumClust^3)$, where $\NumClust$ is the actual number of communities.
We assume that $\NumClust \ll \clustsketchsizedef$, such that the eigendecomposition is still the dominant cost, making the computational complexity of the algorithm $\bigO(\NumNodes^3)$.

\section{Main algorithm: details}
\label{sec:mainalgdetails}

Two helper functions are needed.
The first, $\samplefunc{\Graph}{\SketchSize}$, returns a set of $\SketchSize$ nodes sampled uniformly at random from $\Graph$.
The second infers community membership of nodes in the full graph $\Graph$ based on the sketch community estimates $\ClustSketchSetEst$, as described in \sref{sec:infer}.
This function is defined as follows.

\bigskip
$\inferfunc{\Graph}{\ClustSketchSetEst}$
\begin{enumerate}[topsep=0pt,noitemsep,label={(\arabic*)},ref=\theenumi]

    \item $\ClustEst{\clustiter} \gets \emptyset$ for $\clustiter \in \curly{1,\dots,\NumClustEst}$
    
    \item \AlgCmd{for} each node $\nodeeg \in \Node$ \AlgCmd{do}
          \label{alg:staticinferbegin}
    
    \item \begin{alglevel}{\alglevelone}
          $\clustitertwo \gets \argmax_{\clustiter \in \{1,\dots,\NumClustEst\}} \frac{ \card{ \SetDef{(\nodeeg,\nodeegtwo) \in \Edge}{\nodeegtwo \in \ClustSketchEst{\clustiter}} } }{\card{ \ClustSketchEst{\clustiter} } }$
          \end{alglevel}
          \label{step:staticinfer}
    
    \item \begin{alglevel}{\alglevelone}
          $\ClustEst{\clustitertwo} \gets \ClustEst{\clustitertwo} \cup \{i\}$
          \end{alglevel}
    
    \item \AlgCmd{end for}
          \label{alg:staticinferend}

    \item \AlgCmd{return} partition $\ClustSetEst = \SetDef{\ClustEst{\clustiter}}{\clustiter=1,\dots,\NumClustEst}$
    
\end{enumerate}
\bigskip

\noindent
We use the following definition: given a graph $\Graph=(\Node,\Edge)$ and node set $\Node' \!\subset\! \Node$, the subgraph of $\Graph$ induced by $\Node'$ is denoted $\GraphInducedBy{\Graph}{\Node'}$.
The complete definition of $\mainalg$ follows.

\bigskip

\begin{enumerate}[topsep=0pt,noitemsep,label={(\arabic*)},ref=\theenumi]

    \item $\Graph \gets \GraphSnap{0}$
    \label{step:beginfirst}

    \item $\SketchIdx' \gets \samplefunc{\Graph}{\initialsketchsize}$

    \item $\ClustSnapSetEst{0} \gets \inferfunc{\Graph}{\NBAlg{\GraphInducedBy{\Graph}{\SketchIdx'}}}$
    \label{step:endfirst}

    \item Build sketch set $\SketchIdxSnap{0}$ by sampling $\clustsketchsizedef$ nodes uniformly at random from each community $\ClustIterEst \in \ClustSnapSetEst{0}$. If $\clustsketchsizedef\!>\!\card{\ClustIterEst}$, then include all nodes from $\ClustIterEst$.

    \label{step:proportionfirst}

    \item $\AlgCurrClustCount \gets \card{\ClustSnapSetEst{0}}$.

    \item \AlgCmd{for} $\snaptime=1,2,\dots$ \AlgCmd{do}

    \item \begin{alglevel}{\alglevelone}
          $\Graph \gets \GraphSnap{\snaptime}$
          \end{alglevel}
          \label{alg:mainloopbegin}
          
    \item \begin{alglevel}{\alglevelone}
          $\curly{ \ClustEst{1}, \dots, \ClustEst{\AlgCurrClustCount} }$
          
          \quad$\gets
          \inferfunc{\GraphInducedBy{\Graph}{\NodeSnap{\snaptime} \setminus \NodeBirthEst}}
          {\ClustSnapSketchSetEst\snaptimeminusone}$
          \end{alglevel}
          \label{alg:infer}
          
    \item \begin{alglevel}{\alglevelone}
          \AlgCmd{if} $\NodeBirthEst \ne \emptyset$ \AlgCmd{then}
          \end{alglevel}
          \label{alg:birthbegin}

    \item \begin{alglevel}{\algleveltwo}
          $\GraphBirthEst \gets \GraphInducedBy{\Graph}{\NodeBirthEst}$
          \end{alglevel}

    \item \begin{alglevel}{\algleveltwo}
          $\SketchIdxBirth \gets \samplefunc{\GraphBirthEst} {\clustsketchsizedef}$
          \end{alglevel}

    \item \begin{alglevel}{\algleveltwo}
          $\curly{ \ClustEst{\AlgCurrClustCount+1}, \dots, \ClustEst{\AlgCurrClustCount+\NumClustEst} }$
          
          \quad$\gets \inferfunc{\GraphBirthEst}{\NBAlg{\GraphInducedBy{\GraphBirthEst}{\SketchIdxBirth}}}$
          \end{alglevel}
          
    \item \begin{alglevel}{\algleveltwo}
          $\AlgCurrClustCount \gets \AlgCurrClustCount+\NumClustEst$
          \end{alglevel}
          
    \item \begin{alglevel}{\alglevelone}
          \AlgCmd{end if}
          \end{alglevel}
          \label{alg:birthend}

    \item \begin{alglevel}{\alglevelone}
          \AlgCmd{for} $\clusteg \in \curly{1,\dots,\AlgCurrClustCount}$, where $\card{\ClustEst{\clusteg}} > \SplitSizeThresh$ \AlgCmd{do}
          \end{alglevel}
          \label{alg:splitbegin}
   
    \item \begin{alglevel}{\algleveltwo}
          $\ClustIterSketchEst \gets \ClustEst{\clusteg} \cap \SketchIdxSnap{\snaptimeminusone}$
          \end{alglevel}
    
    \item \begin{alglevel}{\algleveltwo}
          $\GraphSketch \gets \GraphInducedBy{\Graph}{\ClustIterSketchEst}$
          \end{alglevel}
          
    \item \begin{alglevel}{\algleveltwo}
          Let $\Adj$ be the adjacency matrix of $\SubGraph$.
          Calculate the eigenvalues $\NBEig{1},\NBEig{2}, ...$ of $\NBmatrixSurrogate$ [defined in \eqref{eqn:NBsurrogate}].
          \end{alglevel}
          \label{alg:calcspliteigs}
          
    \item \begin{alglevel}{\algleveltwo}
          Calculate $\NumClustEst$ as the maximum value of $i$ such that $\lambda_i > \sqrt{\lambda_1}$.
          \end{alglevel}
          \label{alg:splitdetect}
         
    \item \begin{alglevel}{\algleveltwo}
          \AlgCmd{if} $\NumClustEst > 1$ \AlgCmd{then}
          \end{alglevel}
          \label{alg:splitnumclust}
          
    \item \begin{alglevel}{\alglevelthree}
          $\ClustGraph{\clusteg} \gets \GraphInducedBy{\Graph}{\ClustEst{\clusteg}}$
          \end{alglevel}
          
   \item \begin{alglevel}{\alglevelthree}
          $\ClustSketchSetEst'
          \gets \NBAlg{\GraphInducedBy{\ClustGraph{\clusteg}}{\ClustIterSketchEst}}$

          \end{alglevel}

   \item \begin{alglevel}{\alglevelthree}
          
          $\curly{ \ClustEst{\clusteg}, \ClustEst{\AlgCurrClustCount+1}, \dots, \ClustEst{\AlgCurrClustCount+\NumClustEst-1} }$
          
          \quad$\gets \inferfunc{\ClustGraph{\clusteg}}{\ClustSketchSetEst'}$

          \end{alglevel}
          
    \item \begin{alglevel}{\alglevelthree}
          $\AlgCurrClustCount \gets \AlgCurrClustCount+\NumClustEst-1$
          \end{alglevel}

    \item \begin{alglevel}{\algleveltwo}
          \AlgCmd{end if}
          \end{alglevel}
    
    \item \begin{alglevel}{\alglevelone}
          \AlgCmd{end for}
          \end{alglevel}
          \label{alg:splitend}
    
    \item \begin{alglevel}{\alglevelone}
          \AlgCmd{for} community pairs $\clusteg,\clustegtwo \in \curly{1,\dots,\AlgCurrClustCount}$ \AlgCmd{do}
          \end{alglevel}
          \label{alg:mergebegin}

    \item \begin{alglevel}{\algleveltwo}
          \AlgCmd{if} \eqref{eqn:algdetectmerge} holds \AlgCmd{then}
          \label{alg:mergestep}
          \end{alglevel}

    \item \begin{alglevel}{\alglevelthree}
          $\ClustEst{\clusteg} \gets \ClustEst{\clusteg} \cup \ClustEst{\clustegtwo}$
          \end{alglevel}
          
    \item \begin{alglevel}{\alglevelthree}
          $\ClustEst{\clustegtwo} \gets \emptyset$
          \end{alglevel}  
          
    \item \begin{alglevel}{\algleveltwo}
          \AlgCmd{end if}
          \end{alglevel}
          
    \item \begin{alglevel}{\alglevelone}
          \AlgCmd{end for}
          \end{alglevel}
          \label{alg:mergeend}
          
    \item \begin{alglevel}{\alglevelone}
          $\ClustSnapSetEst{\snaptime} \gets \SetDef{\ClustEst{\clustiter}}{\clustiter=1,\dots,\AlgCurrClustCount}$
          \end{alglevel}
    
    \item \begin{alglevel}{\alglevelone}
          $\NodeDelete \gets \NodeSnap{\snaptimeminusone} \setminus \NodeSnap{\snaptime}$
          \end{alglevel}
          \label{alg:sketchupdatebegin}

    \item \begin{alglevel}{\alglevelone}
          $\SketchIdxSnap{\snaptime} \gets \SketchIdxSnap{\snaptimeminusone} \setminus \NodeDelete$
          \end{alglevel}

    \item \begin{alglevel}{\alglevelone}
          Re-proportion sketch $\SketchIdxSnap{\snaptime}$ such that it contains $\min \{ \clustsketchsizedef, \card{ \ClustIterEst } \}$ nodes from each community $\ClustIterEst \in \ClustSnapSetEst{\snaptime}$.
          \end{alglevel}
          \label{alg:reproportion}

    \item \AlgCmd{end for}
\end{enumerate}

\bigskip

Steps~\ref{step:beginfirst}-\ref{step:endfirst} cluster the first graph snapshot.
This estimate is used to construct a balanced sketch in step~\ref{step:proportionfirst}.
The remainder of the algorithm processes subsequent snapshots.
Step~\ref{alg:infer} re-evaluates the community membership of existing nodes, as well as new nodes joining existing communities.
Steps~\ref{alg:birthbegin}-\ref{alg:birthend} partition the set of newborn communities. Meanwhile, 
Steps~\ref{alg:splitbegin}-\ref{alg:splitend} handle splits within each community.
Only communities with size greater than $\SplitSizeThresh$ are checked, as the spectral estimates become unreliable for small communities.
We set $\SplitSizeThresh\!=\!20$.
Steps~\ref{alg:mergebegin}-\ref{alg:mergeend} handle merges among pairs of communities.
Finally, steps~\ref{alg:sketchupdatebegin}-\ref{alg:reproportion} generate the new sketch.

\section{Derivations for analysis in \sref{sec:analysis}}
\label{sec:proofs}

In this section, we denote a binomial random variable having $\clustsizedef$ trials with probability of success $p$ by $\Bin(n,p)$.
All of the binomial random variables found in this section indicate the number of edges between nodes and/or communities.
Unless otherwise indicated, we assume that the number of edges is sufficient such that $\clustsizedef p \ge 10$, and that the network is sparse enough such that $\clustsizedef (1-p) \ge 10$.
This justifies the use of \asref{as:binomapprox} \cite{devore2012modern}.
We denote a normal random variable with mean $\mu$ and variance $\sigma^2$ by $\Normal(\mu, \sigma^2)$.

\subsection{Initial sketch}

We comment here on the validity of the exact recovery analysis.
The \ac{SBM} model used for analyzing exact recovery in \cite{JMLR:v18:16-480} does not have fixed community sizes, but rather defines a probability distribution over communities. The membership of each node is then sampled from this distribution.
In fact, the initial sketch $\GraphSketchSnap{0}$ adheres to this model, since the probability that a given node in the sketch belongs to community $\clusteg$ is $\frac{\card{\ClustSnap{\clusteg}{0}}}{\initialsketchsize}$.

\subsection{Inferring community membership}

Let $\nummovednodesoutofclustsketch{\clusteg} \!=\! \sum_{\clustitertwo \ne \clusteg} \nummovednodessketch{\clusteg}{\clustitertwo}$ be the total number of sketch nodes moving out of community $\clusteg$, and $\nummovednodesintoclustsketch{\clusteg} \!=\! \sum_{\clustitertwo \ne \clusteg} \nummovednodessketch{\clustitertwo}{\clusteg}$ be the total number of sketch nodes moving into community $\clusteg$.
Then, 
\begin{align} \label{eqn:clustsimdynintraRV}
    \clustsimdyn{\nodeeg}{\clustiter}{\snaptime}
    = \frac{\clustsimdynbinuout + \clustsimdynbinuin}{\clustsketchsizedef}
\end{align}
where
\begin{subequations} \label{eqn:suffinferbinrvs}
\begin{align}
    &\clustsimdynbinuin
    \sim
    \begin{cases}
      \Bin(\clustsketchsizedef-\nummovednodesoutofclustsketch{\clusteg}, \pin), &\clustiter=\clusteg
      \\
      \Bin(\nummovednodessketch{\clustiter}{\clusteg}, \pin),                   &\clustiter \ne \clusteg
    \end{cases}
\\
    &\clustsimdynbinuout
    \sim
    \begin{cases}
      \Bin(\nummovednodesoutofclustsketch{\clusteg}, \pout),                      &\clustiter=\clusteg
      \\
      \Bin(\clustsketchsizedef-\nummovednodessketch{\clustiter}{\clusteg}, \pout), &\clustiter \ne \clusteg
    \end{cases}
\end{align}
\end{subequations}
Since the random variables in \eqref{eqn:suffinferbinrvs} are mutually independent, it follows from 
\asref{as:binomapprox} that
$\clustsimdyn{\nodeeg}{\clusteg}{\snaptime} - \clustsimdyn{\nodeeg}{\clustitertwo}{\snaptime} \sim \Normal(\infersuffnormalmean, \infersuffnormalvar)$
for any $\clustitertwo \ne \clusteg$, where
\begin{align}
\infersuffnormalmean
=& \frac{(\pin-\pout) (\clustsketchsizedef-\nummovednodesoutofclustsketch{\clusteg}-\nummovednodessketch{\clustiter}{\clusteg})}
{\clustsketchsizedef},
\nn\\
\infersuffnormalvar
=& \frac{\pin (1-\pin) + \pout (1-\pout)}{\clustsketchsizedef}
\nn\\
-& \frac{\bracket{\pin (1-\pin) - \pout (1-\pout)}
(\nummovednodesoutofclustsketch{\clusteg}-\nummovednodessketch{\clustiter}{\clusteg})}
{\clustsketchsizedef^2}.
\end{align}
We note that if there are few moving nodes or few edges between communities, then $\clustsimdynbinuin$ and $\clustsimdynbinuout$ may not be well-approximated by a normal random variable.
Nonetheless, in these cases the expected values of $\clustsimdynbinuin$ and $\clustsimdynbinuout$ are small enough that they do not contribute significantly to the bound regardless.

Let $\infersuffnormalboundrv \!\sim\! \Normal(\infersuffnormalmeanbound, \frac{\infersuffnormalvarbound}{\clustsketchsizedef})$.
If \eqref{eqn:suffmoveinfer} holds, then $\infersuffnormalmeanbound \!\ge\! \norminv{1\!-\!\frac{1-\probofsuccess}{\NumClustSnap{\snaptime}-1}} \sqrt{\frac{\infersuffnormalvarbound}{\clustsketchsizedef}}$ and hence
$\prob(\infersuffnormalboundrv\!<\!0) \!\le\! \frac{1-\probofsuccess}{\NumClustSnap{\snaptime}-1}$.
Furthermore, from \asref{as:movesketchexpect}, we have $\infersuffnormalmean \!\ge\! \infersuffnormalmeanbound$ and $\infersuffnormalvar \!\le\! \frac{\infersuffnormalvarbound}{\clustsketchsizedef}$, and thus $\prob(\clustsimdyn{\nodeeg}{\clusteg}{\snaptime}<\clustsimdyn{\nodeeg}{\clustiter}{\snaptime}) \le \prob(\infersuffnormalboundrv<0)$.
Then, by applying the union bound, the probability of successfully inferring the community membership of node $\nodeeg$ is
\begin{align}
    \prob\paren{\bigcap_{\substack{\clustitertwo = 1 \\ \clustitertwo \ne \clusteg}}^{\NumClustSnap{\snaptime}} \clustsimdyn{\nodeeg}{\clusteg}{\snaptime}>\clustsimdyn{\nodeeg}{\clustitertwo}{\snaptime}}
    &= 1-\prob\paren{\bigcup_{\substack{\clustitertwo=1 \\ \clustitertwo \ne \clusteg}}^{\NumClustSnap{\snaptime}} \clustsimdyn{\nodeeg}{\clusteg}{\snaptime}<\clustsimdyn{\nodeeg}{\clustitertwo}{\snaptime}}
    \nn\\
    &\ge 1-\sum_{\substack{\clustitertwo=1 \\ \clustitertwo \ne \clusteg}}^{\NumClustSnap{\snaptime}} \prob\paren{\clustsimdyn{\nodeeg}{\clusteg}{\snaptime}<\clustsimdyn{\nodeeg}{\clustitertwo}{\snaptime}}
    \nn\\
    &= \probofsuccess.
\end{align}

\subsection{Split detection}

Following a similar line of reasoning as in \sref{sec:inititalsketch}, exact recovery is efficiently solvable if
\begin{align} \label{eqn:BoundExactRecoverySplit}
        \frac{2 \clustsketchsizedef}{\ln(2 \clustsketchsizedef)}
        > 2 (\sqrt{\pin} - \sqrt{\pinterclustnotime{\clusteg}{\clustegtwo}})^{-2}\:.
\end{align}
On the other hand, from \eqref{eqn:detectthresh}, the split is asymptotically detectable in the spectrum of $\NBmatrixSurrogate$ if
\begin{align} \label{eqn:detectthreshsketch}
\clustsketchsizedef \paren{\pin - \pinterclustnotime{\clusteg}{\clustegtwo}}^2 >
\pin + \pinterclustnotime{\clusteg}{\clustegtwo},
\end{align}
which is equivalent to
\begin{align} \label{eqn:mergethreshsplitquadratic}
\pinterclustnotime{\clusteg}{\clustegtwo}^2
- \pinterclustnotime{\clusteg}{\clustegtwo}
  \paren{2\pin + \frac{1}{\clustsketchsizedef}}
+ \pin \paren{\pin - \frac{\pin}{\clustsketchsizedef}}
\ge 0.
\end{align}
The bounds in \eqref{eqn:SplitBoundExactRecovery} follow directly by solving \eqref{eqn:BoundExactRecoverySplit} and \eqref{eqn:mergethreshsplitquadratic} in terms of $\pinterclustnotime{\clusteg}{\clustegtwo}$.

\subsection{Merge detection}
Define
$
\maxintraedges=
\NumClustSnap{\snaptime} \clustsketchsizedef(\clustsketchsizedef\!-\!1)/2
$
as the maximum possible number of intracommunity edges at time $\snaptime$.
Then,
\begin{subequations}
\begin{align} 
    \pinest
    &= \frac{\pinestbin}{\maxintraedges}, \label{eqn:pinestRV}
\\
    \pinterclustestnotime{\clusteg}{\clustegtwo}
    &= \frac{\pinterclustestbin{\clusteg}{\clustegtwo}}{(\clustsketchsizedef)^2}
\end{align}
\end{subequations}
where
$
    \pinestbin
    \sim \Bin(\maxintraedges, \pin),
$
and
$
    \pinterclustestbin{\clusteg}{\clustegtwo}
    \sim \Bin((\clustsketchsizedef)^2, \pinterclust{\clusteg}{\clustegtwo}{\snaptime}).
$
Since $\pinestbin$ and $\pinterclustestbin{\clusteg}{\clustegtwo}$ are independent, it follows from \asref{as:binomapprox} that $\pinest - \pinterclustestnotime{\clusteg}{\clustegtwo} \sim \Normal\paren{\mergediffmean, \mergediffsumstd^2}$ and $\pinest + \pinterclustestnotime{\clusteg}{\clustegtwo} \sim \Normal\paren{\mergesummean, \mergediffsumstd^2}$, where
$
\mergesumdiffmean
= \pin \pm \pinterclustnotime{\clusteg}{\clustegtwo}$.

Let $\probofsuccesstwo = 1-\frac{1-\probofsuccess}{2}$.
If condition \eqref{eqn:suffmergeq} holds, then
\begin{align} \label{eqn:suffmergetwo}
    &\pinterclustnotime{\clusteg}{\clustegtwo}^2
    -
    \pinterclustnotime{\clusteg}{\clustegtwo}
    \paren{2\pin+2\norminv{\probofsuccesstwo}\mergediffsumstd+\frac{\DetectThresh^2}{\clustsizedef}}
    \nn\\
    &+
    \paren{\pin + \norminv{\probofsuccesstwo} \mergediffsumstd}^2
    - \DetectThresh^2 \frac{\pin - \norminv{\probofsuccesstwo} \mergediffsumstd}{\clustsizedef}
    < 0
    ,
\end{align}
which in turn implies that
\begin{align} \label{egn:214099}
    \mergediffmean + \norminv{\probofsuccesstwo} \mergediffsumstd
    <
    \DetectThresh \sqrt{\frac{\mergesummean - \norminv{\probofsuccesstwo} \mergediffsumstd}{\clustsizedef}}.
\end{align}
Then, from \eqref{egn:214099} and the union bound, the probability that \eqref{eqn:algdetectmerge} will indicate a split state is
\begin{align}
    &\prob\paren{\pinest - \pinterclustestnotime{\clusteg}{\clustegtwo}
    <
    \DetectThresh \sqrt{\frac{\pinest + \pinterclustestnotime{\clusteg}{\clustegtwo}}{\clustsizedef}}}
    \nn\\
    &\ge \prob\left(\pinest - \pinterclustestnotime{\clusteg}{\clustegtwo}
    < \mergediffmean + \norminv{\probofsuccesstwo} \mergediffsumstd \right.
    \nn\\
    &\qquad \bigcap \left.
         \DetectThresh \sqrt{\frac{\pinest + \pinterclustestnotime{\clusteg}{\clustegtwo}}{\clustsizedef}}
         >
         \DetectThresh \sqrt{\frac{\mergediffmean - \norminv{\probofsuccesstwo} \mergediffsumstd}{\clustsizedef}}
                 \right)
\nn\\
    &\ge 1-\prob\paren{\pinest - \pinterclustestnotime{\clusteg}{\clustegtwo}
    < \mergediffmean + \norminv{\probofsuccesstwo} \mergediffsumstd}
\nn\\
    &\qquad- \prob\paren{\pinest + \pinterclustestnotime{\clusteg}{\clustegtwo}
    > \mergesummean - \norminv{\probofsuccesstwo} \mergediffsumstd}
\nn\\
    &= \probofsuccess
\end{align}
We can derive \eqref{eqn:suffsplitdetq} using a similar argument.

\subsection{Detecting the birth event}

From \eqref{eqn:clustsimdynintraRV} and \eqref{eqn:pinestRV} and under \asref{as:binomapprox} with no moving nodes,
\begin{align}
    \clustsimdyn{\nodeeg}{\clustiter}{\snaptime}
    &\sim \Normal\paren{\pout, \frac{\pout (1-\pout)}{\clustsketchsizedef}},
    1 \le \clustiter \le \NumClustSnap{\snaptime}
\\
    \pinest
    &\sim \Normal\paren{\pin, \frac{\pin (1-\pin)}{\maxintraedges}}
\end{align}

Then $\prob\paren{\pinestlowerbound \le \pinest \le \pinestupperbound} = 1-\birthprob$,
and therefore
\begin{align}
    &\prob\paren{
    \pinest - 3 \sqrt{ \frac{\pinest (1-\pinest)}
                            {\clustsketchsizedef}
                     }
    <
    \pinestlowerbound - 3 \sqrt{\frac{\pinestupperbound(1-\pinestlowerbound)}{\clustsketchsizedef}}
    }
    \nn\\
    &\le 1-\prob\paren{\pinestlowerbound \le \pinest \le \pinestupperbound}
    = \birthprob.
\end{align}
Furthermore, suppose that node $\nodeeg$ is joining community $\clusteg$.
For any community $\clustiter \ne \clusteg$,
\begin{align}
    &\prob\paren{\clustsimdyn{\nodeeg}{\clustiter}{\snaptime}
    \ge \pout + \norminv{1-\birthprob} \sqrt{\frac{\pout(1-\pout)}{\clustsketchsizedef}}}
    = \birthprob.
\end{align}

Condition \eqref{eqn:suffbirth} is equivalent to
\begin{align}
    &\pout
          + \norminv{1-\birthprob}
          \sqrt{\frac{\pout(1-\pout)}{\clustsketchsizedef}}
    \nn\\
    &\qquad \le
    \pinestlowerbound - 3 \sqrt{\frac{\pinestupperbound(1-\pinestlowerbound)}{\clustsketchsizedef}}.
\end{align}
Then, the probability that the condition inside \eqref{eqn:birthnodenodeest} will fail for a particular $\clustiter$ is
\begin{align}
    &\prob\paren{
    \clustsimdyn{\nodeeg}{\clustiter}{\snaptime}
    \ge
    \pinest - 3 \birthstdest{\clustiter}
    }
\nn\\
    &= \prob\paren{
    \clustsimdyn{\nodeeg}{\clustiter}{\snaptime}
    \ge
    \pinest - 3 \sqrt{ \frac{\pinest (1-\pinest) }{\clustsketchsizedef}}
    }
\nn\\
    &\le \prob\left(
    \clustsimdyn{\nodeeg}{\clustiter}{\snaptime}
    \ge
    \pout
    + \norminv{1-\birthprob}
    \sqrt{\frac{\pout(1-\pout)}{\clustsketchsizedef}}
    \right.
\nn\\
    &\qquad\bigcup
    \left.
    \pinest - 3 \sqrt{ \frac{\pinest (1-\pinest)}
                            {\clustsketchsizedef}
                     }
    <
    \pinestlowerbound - 3 \sqrt{\frac{\pinestupperbound(1-\pinestlowerbound)}{\clustsketchsizedef}}
    \right)
\nn\\
    &\le \prob\paren{
    \clustsimdyn{\nodeeg}{\clustiter}{\snaptime}
    \ge
    \pout
    + \norminv{1-\birthprob}
    \sqrt{\frac{\pout(1-\pout)}{\clustsketchsizedef}}
    }
\nn\\
    &\quad +
    \prob\paren{
    \pinest - 3 \sqrt{ \frac{\pinest (1-\pinest)}
                            {\clustsketchsizedef}
                     }
    <
    \pinestlowerbound - 3 \sqrt{\frac{\pinestupperbound(1-\pinestlowerbound)}{\clustsketchsizedef}}
    }
\nn\\
    &= 2\birthprob.
\end{align}

Consequently, the probability that the condition will hold for all communities is
\begin{align}
    &\prob\paren{
        \bigcap_{\substack{v=1 \\ v \ne \clusteg}}^\NumClust \clustsimdyn{\nodeeg}{\clustiter}{\snaptime} + 3 \birthstdest{\clustiter} < \pinest
    }
\nn\\
    &= 1-\prob\paren{\bigcup_{\substack{v=1 \\ v \ne \clusteg}}^\NumClust
        \clustsimdyn{\nodeeg}{\clustiter}{\snaptime} + 3 \birthstdest{\clustiter} \ge \pinest}
\nn\\
    &\ge 1-\sum_{\substack{v=1 \\ v \ne \clusteg}}^\NumClust \prob\paren{\clustsimdyn{\nodeeg}{\clustiter}{\snaptime} + 3 \birthstdest{\clustiter} \ge \pinest}
\nn\\
    &= 1 - 2 \paren{\NumClustSnap{\snaptime}-1} \birthprob
    = \probofsuccess.
\end{align}

\section{Numerical results: details of algorithms used in comparison}
\label{sec:comparisonalgorithms}

For the algorithm of Yang \textit{et al.} \Yang/, we use the same tuning strategy as in the experimental results section of \cite{Yang2011}.
Specifically, we use the same temperature and iteration sequences, with $\alpha=0.8,\beta=0.5,\gamma=1,\mu_{kk}=10$.
We run five instances of the algorithm with
(1) $\alpha_{kk}=1, \beta_{kl}=1$,
(2) $\alpha_{kk}=5, \beta_{kl}=1$,
(3) $\alpha_{kk}=10, \beta_{kl}=1$,
(4) $\alpha_{kk}=10^2, \beta_{kl}=10$,
(5) $\alpha_{kk}=10^4, \beta_{kl}=10$,
and then take the community assignments among the five trials yielding the highest average modularity (modularity is defined as in \cite{Yang2011}).
For the algorithm of Dinh \textit{et al.} \Dinh/, we use the CNM algorithm \cite{PhysRevE.70.066111} for the static clustering step, as in \cite{5403845}.
When running the \acs{ESPRA} algorithm, we use the same parameters as used in the experimental results of \cite{Wang_2017}: $\alpha=0.8, \beta=0.5$.
The algorithm \StdInd/ applies $\NBAlgnoparams$ to each snapshot to obtain community estimates $\curly{\ClustEstUnmatched{1},\dots,\ClustEstUnmatched{\NumClustEst}}$.
To provide continuity in the community assignments of the nodes, community $\clusteg$ in each snapshot at time $\snaptime\!>\!0$ is matched to the community at time $\snaptimeminusone$ having the largest overlap according to the Jaccard coefficient.
Specifically, for each community $\clusteg$, we set the new estimate as $\ClustSnapEst{\clusteg}{\snaptime} \!=\! \ClustEstUnmatched{\clustegtwo}$ where
$
\clustegtwo = \argmax_{1 \le \clustiter \le \NumClustEst} \frac{\card{ \ClustEstUnmatched{\clusteg} \cap \ClustSnapEst{\clustiter}{\snaptimeminusone} }}
                          {\card{ \ClustEstUnmatched{\clusteg} \cup \ClustSnapEst{\clustiter}{\snaptimeminusone} }}.
$

\bibliographystyle{IEEEtran}

\bibliography{main}

\end{document}